\documentclass[twocolumn,showpacs,showkeys,preprintnumbers,superscriptaddress,amsmath,floatfix,amssymb,secnumarabic,nofootinbib]{revtex4-1}
\usepackage[colorlinks=true]{hyperref}
\usepackage{graphicx}
\usepackage{dblfloatfix}    % To enable figures at the bottom of page
\usepackage{placeins}
\usepackage{braket}
\usepackage{amsmath}
\usepackage{epsfig}
\usepackage{float}
\usepackage{nicefrac}
\usepackage{xcolor}
\usepackage{color}
\usepackage{bm}
\usepackage{subfigure}
\usepackage{chngcntr}
\usepackage{ulem}

\def\({\left(}
\def\){\right)}
\def\[{\left[}
\def\]{\right]}

\newcommand{\beq} {\begin{eqnarray}}
\newcommand{\eeq} {\end{eqnarray}}

\begin{document}
\sloppy

\title{Flat  band projections: Sign problem mapping for frustrated spin systems}

\author{Maksim~Ulybyshev}
\email{maksim.ulybyshev@uni-wuerzburg.de}
\affiliation{Institut f\"ur Theoretische Physik und Astrophysik, Universit\"at W\"urzburg, 97074 W\"urzburg, Germany}

\author{Anika~G\"otz}
\email{anika.goetz@uni-wuerzburg.de}
\affiliation{Institut f\"ur Theoretische Physik und Astrophysik, Universit\"at W\"urzburg, 97074 W\"urzburg, Germany}

\author{Fakher~Assaad}
\email{fakher.assaad@uni-wuerzburg.de}
\affiliation{Institut f\"ur Theoretische Physik und Astrophysik, Universit\"at W\"urzburg, 97074 W\"urzburg, Germany}
\affiliation{W\"urzburg-Dresden Cluster of Excellence ct.qmat, Am Hubland, 97074 W\"urzburg, Germany}

\begin{abstract}
   
Projection of  the  Coulomb potential  onto flat  bands  paves  the  way   to  design  various interactions in the particle-hole  and particle-particle  channels.  
Here  we pose  the  question  if  we  can use  this mapping to  overcome  the   negative sign  problem for  the simplest  possible  frustrated  spin system   consisting  a  trimer of  spins-1/2   coupled  with an  antiferromagnetic  exchange interaction.    While  the  answer  is  negative,  we   show  that    we  can map the sign problem for frustrated spin systems onto a  problem where we need to simulate particle-hole symmetric systems but with long-ranged Coulomb interactions prevailing over short-ranged ones.  While the latter systems are currently not accessible to   auxiliary  field  determinant  quantum  Monte  Carlo,  this mapping motivates algorithmic  development   that  may  overcome  this issue. 

\end{abstract}

\maketitle

\section*{\label{sec:Intro}Introduction}

The sign problem is one of the major obstacles in the  field of Quantum Monte Carlo  (QMC) calculations   \cite{Troyer05}. Only partial success was reached in solving it, with various approaches usually performing well only for particular systems  \cite{PhysRevB.89.111101, Zoltan_Fodor_2007, PhysRevD.105.054501}.
While  many  interesting    energy  scales in spin   \cite{Alet16,Honecker16, SatoT20_1}  and  charged
  \cite{Huang19,Ulybyshev19,ZhangX_23}  systems  can  be 
 achieved  by finding  optimal 
formulations   that  maximize     
the   average  sign  and  then confronting  it  with   efficient  algorithms,   it  remains  the  major  drawback    of  
the   stochastic  approach  to  correlated  electron  systems.  There  is  a  notion  that  the  sign  problem   is  tied  to  
Hamiltonian classes  or  phases of  quantum  matter \cite{Grossman23}.   For  instance,   the  physics  of  local  magnetic  moments,  generated  by a   
screened   Coulomb   repulsion  in a   translationally  invariant  metallic   system,   remains  very  challenging.    One  instance  of  this  class  is  the 
doped  repulsive  Hubbard  model.    Possible  solutions  to  this  class  of  problems  were  put  forward  in   Ref.~\cite{Corney04},    in terms  of  
a   sign problem  free diffusion  Monte  Carlo   approach  in the space  of   Gaussian operators.  However,  the   outcome was 
that  the  approach  was  flawed   \cite{Assaad05,Corboz08}  most probably  due  to  fat  tailed    distributions.  In this  sense,  the  sign   problem   was  mapped  onto 
another  problem  of   equivalent  complexity.   

 	Mapping  the  sign  problem  of  one  system onto  another is  important since  it  highlights  the   impact    of potential  solutions  to  the 
sign  problem  for  certain Hamiltonians.    Here  we   map  the Monte Carlo  simulations  of   frustrated spin systems onto  the  problem  of 
simulating  particle-hole symmetric  charged  systems   with long-ranged Coulomb interactions prevailing over short-ranged ones. 
%which prevents reliable simulations of e.g. spin frustrated systems or Hubbard model away of half-filling. 
To  be  specific,  we propose a lattice model with a flat band and long-range charge-charge interactions, which hosts a spin frustrated system as an effective model for the flat band. This model provides a mapping of the sign problem for spin frustrated system onto the problem of simulating some system, where long-range interactions dominate over the  Hubbard repulsion. The latter systems are not accessible for at least determinantal QMC either, so the solution of the sign problem remains elusive. However,  progress in algorithms might improve the situation in the  future, and the mapping might become useful in this case. In addition to that, the mapping can serve as a proof of equivalence of complexity of the sign problem for the spin frustrated systems and  systems dominated by long-range charge-charge interactions. 

The paper is organized as follows: in the first section, we describe the geometry and free electron spectrum of the proposed lattice model. We also give an account of electron-electron interactions and describe the projection onto the flat band. 
The second section is devoted to  the  verification of the projection algorithm using QMC data. The third section describes the model 
which leads to the effective Hamiltonian with frustrated spin-spin interaction in the flat band. The appendix contains a brief excursion 
into the possibility of using the same technique of flat band projection to generate  effective models with dominant long-range charge-charge interactions.

\section{\label{sec:ModelProj}Lattice model and flat band projection}
\subsection{\label{subsec:ModelGeometry}Lattice geometry}

The design of the lattice model is defined by two main considerations: first, we need well-defined nearly-flat band, which naturally enhances the interaction effects; second, we need localized spins inside this flat band. 

\begin{figure}[]
   \centering
 \includegraphics[width=0.3\textwidth , angle=0]{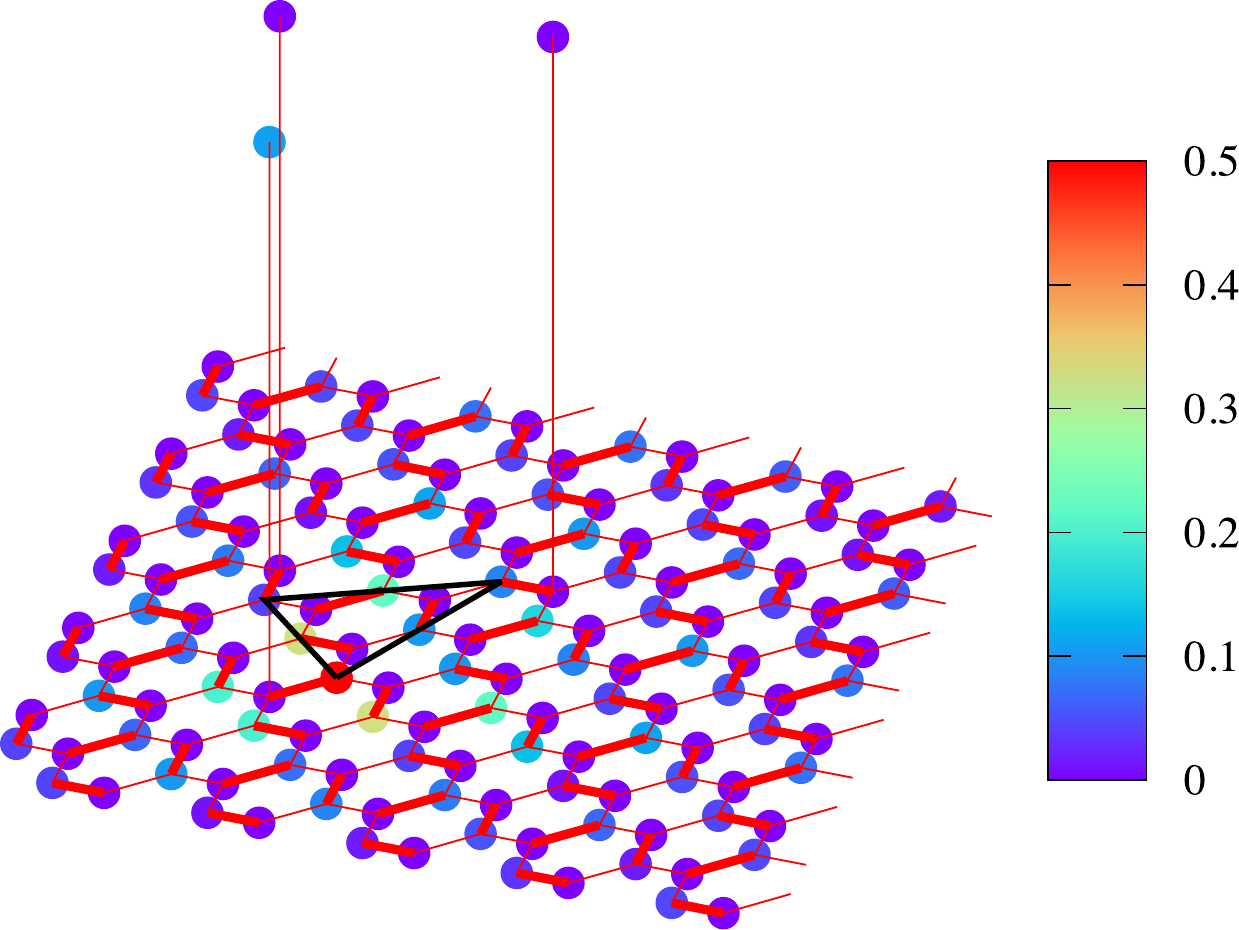}
         \caption{Scheme of the lattice model with three additional sites ("adatoms"). Red lines represent hoppings, Kekule pattern is shown with bold and thin lines.  Bold ones  correspond to larger hopping $t=1.3$ and the thin lines correspond to smaller hopping $t=1.0$. Hoppings to additional sites are equal to $t'=10.0$.  Black lines connect the sites involved in charge-charge interactions. Color scale corresponds to the modulus of the Wannier wavefunction, concentrated around one of the adatoms. }
   \label{fig:system1_scheme}
\end{figure}

The first argument leads to the system, where the flat band is embedded inside the gap between conduction and valence bands. Thus we take the hexagonal lattice with nearest-neighbour hoppings following Kekule pattern to create a mass gap. Next, we introduce additional sites similar to adatoms on top of carbon atoms in graphene. Each of these additional sites features a single hopping to the closest regular site of the hexagonal lattice. The kinetic part of the Hamiltonian can be written as follows:
\begin{eqnarray}
\hat H_K= \sum_{<x,y>\in \mathcal{H} } t_{<x,y>} ( \hat a^\dag_{x, \sigma} \hat a_{y, \sigma} + \hat a^\dag_{y, \sigma} \hat a_{x, \sigma} ) + \nonumber \\  \sum_{<z,y>, z\in\mathcal{A}} t' ( \hat a^\dag_{z, \sigma} \hat a_{y, \sigma} + \hat a^\dag_{y, \sigma} \hat a_{z, \sigma} ),
    \label{eq:kin_hamiltonian}
\end{eqnarray}
where $<x,y>$ are nearest neighbour sites, $\mathcal{H}$ is the set of sites in regular hexagonal lattice, $\mathcal{A}$ is the set of additional sites. $\hat  a_{y, \sigma}$ are annihilation operators for electron with spin $\sigma=\uparrow, \downarrow$ on site $x$. Summation over spin indexes $\sigma$ is implied. Hoppings inside the hexagonal lattice $t_{<x,y>}$ follow the Kekule pattern.  The spatial modulation  of the hoppings is illustrated  in the 
Fig.~\ref{fig:system1_scheme}.  Note, that despite the presence of additional sites, the lattice remains bipartite. Thus, due to particle-hole symmetry,  one isolated adatom creates one additional state in the spectrum of single-particle Hamiltonian, and this state is located at exactly zero energy.  If we have several adatoms, all connected to the sites of hexagonal substrate located at the same sublattice,  particle-hole symmetry prohibits direct hoppings between the states in the flat band, thus creating a perfectly flat band.  If adatoms are connected to different sublattices, some dispersion appears in the flat band but the general particle-hole symmetry is still preserved. We are interested in perfectly flat band, thus we will consider mostly the former case, as  reflected in the example of Fig.~\ref{fig:system1_scheme}.

The hopping $t'$ controls the properties of the wavefunctions inside the flat band. If $t'\ll t_{<x,y>}$, these wavefunctions become completely spatially separated from the states in the conduction and valence band: flat band states are concentrated mostly at adatoms, where conduction and valence band states have almost zero electronic density. It will be evident from the projection algorithm, that we need some non-trivial matrix elements of the interaction Hamiltonian $\hat H_U$ between valence (conduction) bands and the flat band, hence this complete spatial separation is quite disadvantageous. Thus we choose to work in opposite limit, when $t'\gg t_{<x,y>}$. In this case, the wavefunctions of the flat band states lie mostly within the underlying hexagonal lattice. 

If $\hat H_U$ includes only   the Hubbard interaction, a  Lieb  
theorem \cite{Lieb89}  prevents the formation of any non-trivial spin state. In  particular,   since  we  place   the  adatoms on the  same sub-lattice,  the  ground  state  total spin   is  given by   $N_{\text{adatoms}}/2$ corresponding  to  a  ferromagnetic  state.
Thus we need to include some long-range interactions via the expression
\begin{eqnarray}
\hat H_U= \frac{1}{2} \sum_{xy \in \mathcal{H} } U_{xy} \hat q_{x} \hat q_{y},
    \label{eq:int_hamiltonian}
\end{eqnarray}
where $U_{xy}$ is a matrix of charge-charge interactions, and $\hat q^\dag_{x} =     \sum_{\sigma} \left(  \hat{a}^{\dag}_{x,\sigma} \hat{a}^{}_{x,\sigma}  -  \frac{1}{2} \right) $ is the charge operator at site $x$. For our purposes, it is not even needed for all sites to participate in the interaction, it is enough to just include some sites concentrated around adatoms, where the electronic density of the flat band states peaks. Existing QMC algorithms impose some limitations on the matrix $U$: for  repulsive interactions,  it  has  to be positive  definite \cite{Ulybyshev2013,Hohenadler14}.
We can also perform simulations in the case when all eigenvalues are non-positive (e.g. attractive Hubbard model). However, existing QMC algorithms fail if there is a mixture of positive and negative eigenvalues in the $U$ matrix.

\subsection{\label{sec:Proj}Projection method}

We start from the diagonalization of the free Hamiltonian $\hat H_K$, which gives us $N_L$ states in lower (valence) band, $N_U$ states in upper (conduction) band and $N_0$ states in the flat band, where $N_0$ is equal to the number of adatoms. We introduce the field operators for these states
\begin{eqnarray}
 \hat \psi_{i, \sigma}=\sum_x \psi_i(x) \hat a_{x, \sigma},
 \label{eq:field_op}
\end{eqnarray}
where $\psi_i(x)$   is the single-electron wavefunction of the $i$-th eigenvalue of the free Hamiltonian $\hat H_K$, including both upper, lower, and flat band. In the case of degenerate energy levels inside the flat band, we choose maximally localized Wannier functions \cite{RevModPhys.84.1419}. Example of one such wave function, localized near one of the adatoms, is shown in Fig.~\ref{fig:system1_scheme} using the color of the lattice sites.

Using this single-particle basis, a general vector in the Hilbert space for 3-band system can be written as follows: 
\begin{eqnarray}
    | \Psi \rangle = | n^{L}_{1, \sigma}, n^{L}_{2, \sigma} ... n^{L}_{N_L, \sigma}; n^{0}_{1, \sigma} ... n^{0}_{N_0, \sigma};  n^{U}_{1, \sigma},  ... n^{U}_{N_U, \sigma} \rangle.
    \label{eq:wf_general}
\end{eqnarray}
$\{n^{U}\}$, $\{n^{L}\}$ and $\{n^{0}\}$ are occupation numbers (eigenvalues of the particle number  operators  $\hat \psi^{\dagger}_{i, \sigma}   \hat \psi_{i, \sigma} $) for both spin projections for energy levels in upper, lower and flat band correspondingly. Of course, the system can have more than 3 bands, but for simplicity we joined all bands below the flat band in one "lower" band and all bands above the flat band in the "upper" one. 

Since the flat band is energetically separated from upper and lower bands (we will give a numerical estimation for the degree of separation needed below), we develop the effective 
theory for the flat band by projecting into the subspace $\mathcal{P}$, where the lower band is completely filled and the upper band is empty.  
\begin{eqnarray}
    | \Psi_P \rangle = \hat P | \Psi \rangle = | 1, ... 1; n^{0}_{1, \sigma} ... n^{0}_{N_0, \sigma};  0,  ... 0 \rangle.
    \label{eq:wf_projected}
\end{eqnarray}
We stress that the symmetries of  the original  model such  as   global  spin  and  charge  invariance  carry  over  to  the  effective  model  defined in 
the  projected  Hilbert  space.

A general expression for the effective Hamiltonian inside the projected subspace is written as follows:
\begin{eqnarray}
    \hat H_{eff.}(E)=\hat P \hat H \hat P + \hat P \hat H \hat R \frac{1}{E-\hat R \hat H \hat R} \hat R \hat H \hat P,
    \label{eq:eff_ham_base}
\end{eqnarray}
where $\hat H = \hat H_K + \hat H_U$,  $\hat R = I - \hat P$, and $E$ is the energy.
This expression can be derived e.g. using the Schur complement starting from the equation for the eigenvalues of the initial Hamiltonian in the full Hilbert space: 
\begin{eqnarray}
& & \det(  \hat{H} -  E    )    =   \det 
\begin{pmatrix}
\hat P \hat H \hat P - E &   \hat P \hat H \hat  R   \\
\hat R \hat H \hat  P    & \hat R \hat H \hat  R  - E
\end{pmatrix}  =  \nonumber   \\ 
& & 
\det(   \hat R \hat H \hat  R -  E    )   \\
& &  \times \det(    P \hat H \hat P - E    -   \hat P \hat H \hat  R   \frac{1}{ \hat R \hat H \hat  R  - E }  \hat R \hat H \hat  P   ).   \nonumber 
\end{eqnarray} 
Since the system features a gap, the desired subspace with dynamics concentrated in the flat band only forms a low energy sector of the full set of eigenvalues. Thus we can assume that the term  $ \det(   \hat R \hat H \hat  R -  E    )  $, which features the projection outside of the  $\mathcal{P}$ subspace  never  vanishes.  Hence we can consider only the second term, leading exactly to the expression \ref{eq:eff_ham_base}.

Since the expression for the effective Hamiltonian includes the energy, we will need to solve the equation for the eigenvalues
\begin{eqnarray}
\det \left( {H_{eff}}_{ij} (E) - \delta_{ij} E \right)=0
    \label{eq:det_eq}
\end{eqnarray}
self-consistently including the dependence on energy inside the matrix of the effective Hamiltonian ${H_{eff}}_{ij} (E)$.

 \begin{figure}[]
   \centering
   \subfigure[]{\label{fig:diagramV1}\includegraphics[width=0.3\textwidth , angle=0]{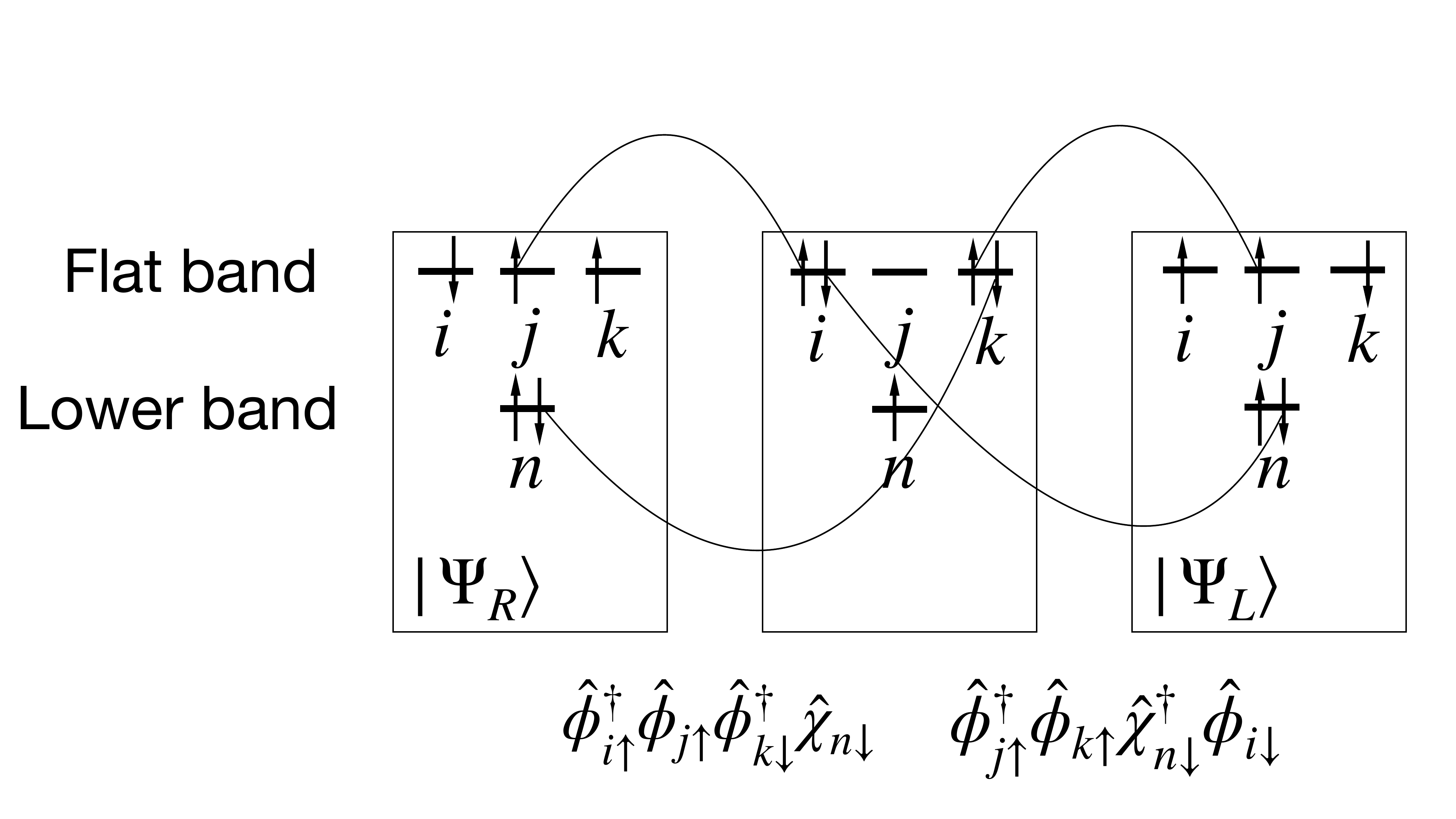}}
   \subfigure[]{ \label{fig:diagramV2}\includegraphics[width=0.3\textwidth , angle=0]{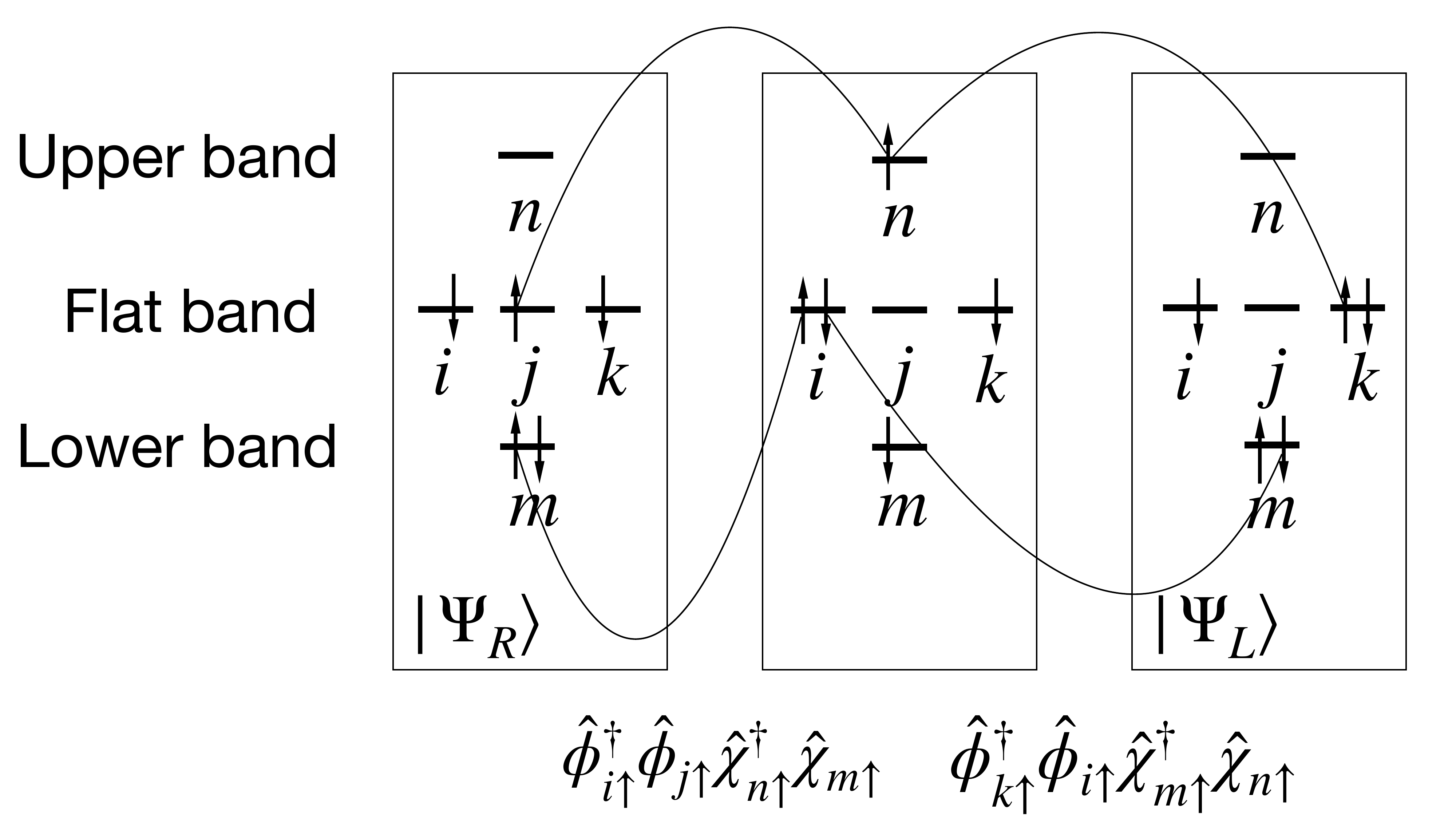}}
    \subfigure[]{ \label{fig:diagramV3}\includegraphics[width=0.3\textwidth , angle=0]{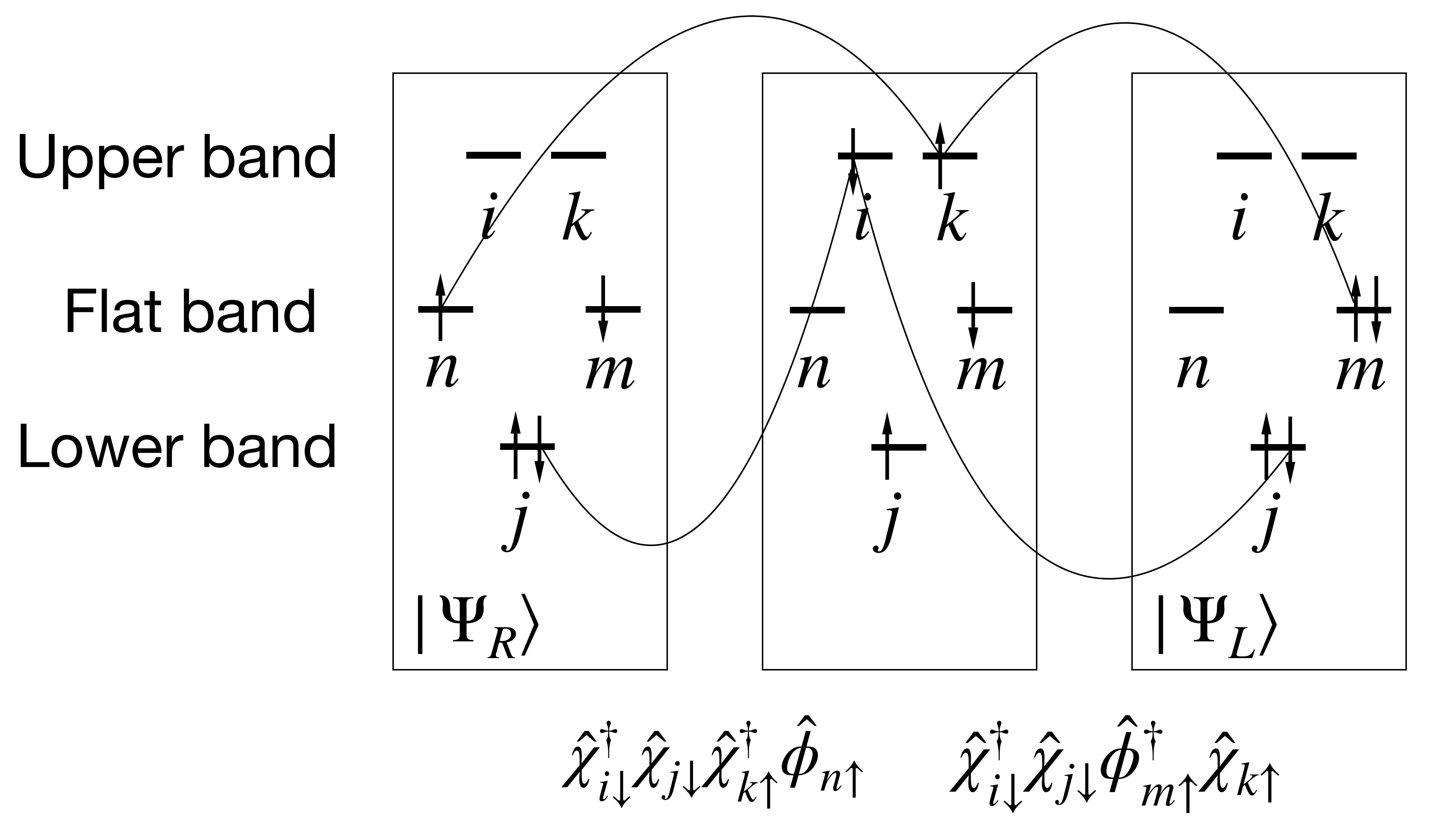}}
         \caption{Examples of diagrams for matrix elements $\langle \Psi_L | \hat H_{eff} |\Psi_R \rangle$, following from the second term in \ref{eq:eff_ham1} under the assumption that only diagonal operators in the basis \ref{eq:wf_general} are left in the denominator. Diagram (a) corresponds to some particular terms in $\hat V^{(1,1)}$ and  $\hat V^{(1,2)}$; (b) corresponds to $\hat V^{(2,1)}$; (c) corresponds to $\hat V^{(3,1)}$ and  $\hat V^{(3,2)}$. Operator terms shown in the diagrams are written below the arrows which indicate the transitions caused by these terms. }
         \label{fig:diagramV}
\end{figure}

Now, we rewrite the effective Hamiltonian \ref{eq:eff_ham_base} using the field operators  \ref{eq:field_op}.
 First, we separate the part of Hamiltonian which can not transfer the states outside subspace $\mathcal{P}$: 
\begin{eqnarray}
\hat H=\hat H_\phi + \hat V, \label{eq:h_psi_V} \\
    \hat P \hat H_\phi \hat R =0.
\end{eqnarray}
For simplicity, we will denote the field operators for the flat band modes as $\hat \phi_{i, \sigma}$, and field operators for upper and lower bands as $\hat \chi_{i, \sigma}$. 
Thus the operator $\hat H_\phi$ in \ref{eq:h_psi_V} can be written in the following general form 
\begin{eqnarray}
\hat H_\phi = \sum_{ijkn} H^{(1)}_{ijkn} \hat \phi^\dag_{i, \sigma} \hat \phi_{j, \sigma} \hat \phi^\dag_{k, \sigma'} \hat \phi_{n, \sigma'} + \sum_{ij} H^{(2)}_{ij} \hat \phi^\dag_{i, \sigma} \hat \phi_{j, \sigma}  \nonumber \\ 
+ \sum_{ijk} H^{(3)}_{ijk} \hat \phi^\dag_{i, \sigma} \hat \phi_{j, \sigma} \hat \chi^\dag_{k, \sigma'} \hat \chi_{k, \sigma'} + \sum_{i} H^{(4)}_{i} \hat \chi^\dag_{i, \sigma} \hat \chi_{i, \sigma}   \label{eq:H_psi_V_def} \\
+ \sum_{ij} H^{(5)}_{ij} \hat \chi^\dag_{i, \sigma} \hat \chi_{i, \sigma} \hat \chi^\dag_{j, \sigma'} \hat \chi_{j, \sigma'}. \nonumber 
\end{eqnarray}
As in \ref{eq:kin_hamiltonian}, summation over spin indexes is implied.
$H^{(i)}$ coefficients can be easily computed after substitution of the field operators \ref{eq:field_op} in Hamiltonian $\hat H$. For example,
\begin{eqnarray}
H^{(1)}_{ijkn}=\frac{1}{2}\sum_{x,y} U_{x,y} \left( \phi_i (y) \phi^*_j (y) \phi_k (x) \phi^*_n (x) \nonumber \right. \\ + \left. \phi_i (x) \phi^*_j (x) \phi_k (y) \phi^*_n (y) \right)
 \label{eq:H1_coef_example}
\end{eqnarray}

 $\hat H_\phi$ is constructed using the principle, that any expression involving $\hat \chi$ operators allowed in it should consist only from particle number operators $\hat n^{(\chi)}_{i, \sigma}=\hat \chi^\dag_{i, \sigma} \hat \chi_{i, \sigma}$ in various combinations. Thus  $\hat H_\phi$  can be further simplified, since $\hat P$   $ \chi^\dag_{i, \sigma} \hat \chi_{j, \sigma'}  =  \delta_{i,j} \delta_{\sigma,\sigma'}  \epsilon $   with 
$\epsilon  = 0,1$ and all particle number operators $\hat n^{(\chi)}_{i, \sigma}$ can be simply substituted by corresponding occupation numbers.  
Examples  of   $\hat H_\phi$    include    effective  models  for  twisted   bilayer graphene  \cite{Bistritzer11,Khalaf21,Hofmann22,ZhangX_23}  as    well as  landau  level  projections \cite{WangZ20,WeiZ23}. 
 Here  we  go  beyond  this  approximation  and  include   corrections  occurring  from    the  high energy  bands.

The  operator $\hat V$ contains all remaining terms in the Hamiltonian $\hat H$:
\begin{eqnarray}
\hat V & = & \hat V^{(1)} + \hat V^{(2)} + \hat V^{(3)} \\
& +  & \sum_{i\neq j, k\neq n} V^{(4)}_{ijkn} \hat \chi^\dag_{i, \sigma} \hat \chi_{j, \sigma} \hat \chi^\dag_{k, \sigma'} \hat \chi_{n, \sigma'} 
+ \sum_{i \neq j} V^{(5)}_{ij} \hat \chi^\dag_{i, \sigma} \hat \chi_{j, \sigma}  \nonumber 
\label{eq:V_def}
\end{eqnarray}
where $\hat V^{i}, i=1,2,3$ operators correspond to the i-particle excitations to upper (lower) band: 
\begin{eqnarray}
\hat V^{(1)} = \sum_{ijkn} V^{(1,1)}_{ijkn} \hat \phi^\dag_{i, \sigma} \hat \phi_{j, \sigma} \hat \phi^\dag_{k, \sigma'} \hat \chi_{n, \sigma'} +  \nonumber \\ 
 \sum_{ijkn} V^{(1,2)}_{ijkn} \hat \phi^\dag_{i, \sigma} \hat \phi_{j, \sigma} \hat \chi^\dag_{k, \sigma'} \hat \phi_{n, \sigma'} +  \\ 
 + \sum_{ij} V^{(1,3)}_{ij} \hat \phi^\dag_{i, \sigma} \hat \chi_{j, \sigma} + 
\sum_{ij} V^{(1,4)}_{ij} \hat \chi^\dag_{i, \sigma} \hat \phi_{j, \sigma} \nonumber \\
 + \sum_{ijk} V^{(1,5)}_{ijkn} \hat \phi^\dag_{i, \sigma} \hat \chi_{j, \sigma} \hat \chi^\dag_{k, \sigma'} \hat \chi_{k, \sigma'}  \nonumber \\
  + \sum_{ijk} V^{(1,6)}_{ijkn} \hat \chi^\dag_{i, \sigma} \hat \phi_{j, \sigma} \hat \chi^\dag_{k, \sigma'} \hat \chi_{k, \sigma'}  \nonumber 
    \label{eq:V1_def}
\end{eqnarray}
\begin{eqnarray}
\hat V^{(2)} = \sum_{ij,k \neq n} V^{(2,1)}_{ij,kn} \hat \phi^\dag_{i, \sigma} \hat \phi_{j, \sigma} \hat \chi^\dag_{k, \sigma'} \hat \chi_{n, \sigma'} +  \nonumber \\ 
 \sum_{ijkn} V^{(2,2)}_{ijkn} \hat \phi^\dag_{i, \sigma} \hat \chi_{j, \sigma} \hat \phi^\dag_{k, \sigma'} \hat \chi_{n, \sigma'} +   \nonumber \\ 
  \sum_{ijkn} V^{(2,3)}_{ijkn} \hat \chi^\dag_{i, \sigma} \hat \phi_{j, \sigma} \hat \chi^\dag_{k, \sigma'} \hat \phi_{n, \sigma'} +  \\ 
    \sum_{ijkn} V^{(2,4)}_{ijkn} \hat \phi^\dag_{i, \sigma} \hat \chi_{j, \sigma} \hat \chi^\dag_{k, \sigma'} \hat \phi_{n, \sigma'}  \nonumber
     \label{eq:V2_def}
\end{eqnarray}
\begin{eqnarray}
\hat V^{(3)} = \sum_{i \neq j, kn} V^{(3,1)}_{ijkn} \hat \chi^\dag_{i, \sigma} \hat \chi_{j, \sigma} \hat \chi^\dag_{k, \sigma'} \hat \phi_{n, \sigma'} +  \nonumber \\ 
 \sum_{i \neq j,kn} V^{(3,2)}_{ijkn} \hat \chi^\dag_{i, \sigma} \hat \chi_{j, \sigma} \hat \phi^\dag_{k, \sigma'} \hat \chi_{n, \sigma'} 
    \label{eq:V3_def}
\end{eqnarray}
Coefficients $V^{(i)}$ again can be easily obtained after applying the transformation \ref{eq:field_op} to $\hat a_{x, \sigma}$ operators in $\hat H$. 

Now, the effective Hamiltonian  \ref{eq:eff_ham_base} can be rewritten as
\begin{eqnarray}
     \hat H_{eff.}(E)=\hat H_\phi + \hat P \hat V \hat R \frac{1}{E-\hat R (\hat H_\phi + \hat V) \hat R} \hat R \hat V \hat P.
    \label{eq:eff_ham1}
\end{eqnarray}
This expression still gives exact values of energy, since no approximations were made. However, the second term  includes the inversion of the interaction Hamiltonian  $\hat V$, thus it can not be evaluated exactly. Starting from this point we make the approximations assuming that the gap in one particle spectrum  between upper and lower band is larger than matrix elements of the interaction Hamiltonian. The  intermediate states $| \Psi_I \rangle =  \hat R \hat V \hat P | \Psi \rangle $ in the second term in \ref{eq:eff_ham1} have at least one excitation in upper or lower band. Thus, the denominator in \ref{eq:eff_ham1} is dominated by the kinetic energy of these excitations given by
\begin{eqnarray}
\hat H^{(4)}=\sum_{i} H^{(4)}_{i} \hat \chi^\dag_{i, \sigma} \hat \chi_{i, \sigma}
 \label{eq:H4}
\end{eqnarray}
term in $\hat H_\phi $ Hamiltonian, with all other terms only giving subleading contributions.  In addition to the large kinetic term, we also assume the presence of relatively large Hubbard-type interactions, leading to the localization of the states especially in the flat band.  Consequently, we retain additional terms in the denominator of \ref{eq:eff_ham1}. These are essentially all terms, which are diagonal in the basis \ref{eq:wf_general}. In the end the following operators are left in the denominator: $\hat H^{(4)}, \, \hat H^{(5)},  \hat {\tilde H}^{(1)}, \hat {\tilde H}^{(2)}, \hat {\tilde H}^{(3)}$. $\hat H^{(4)}$ is defined in \ref{eq:H4}. The remaining four operators are defined as
\begin{eqnarray}
\hat {H}^{(5)} = \sum_{ij} H^{(5)}_{ij} \hat \chi^\dag_{i, \sigma} \hat \chi_{i, \sigma} \hat \chi^\dag_{j, \sigma'} \hat \chi_{j, \sigma'}. \nonumber \\
\hat {\tilde H}^{(1)} = \sum_{ij} H^{(1)}_{iijj} \hat \phi^\dag_{i, \sigma} \hat \phi_{i, \sigma} \hat \phi^\dag_{j, \sigma'} \hat \phi_{j, \sigma'},\nonumber \\
\hat {\tilde H}^{(2)} = \sum_{i} H^{(2)}_{ii} \hat \phi^\dag_{i, \sigma} \hat \phi_{i, \sigma},   \\
\hat {\tilde H}^{(3)} = \sum_{ik} H^{(3)}_{iik} \hat \phi^\dag_{i, \sigma} \hat \phi_{i, \sigma} \hat \chi^\dag_{k, \sigma'} \hat \chi_{k,\sigma'}. \nonumber
\label{eq:H_psi_tilde} 
\end{eqnarray}
Under these assumptions, all terms in  $\hat V^{(i)}$ operators can be depicted as parts of diagrams for the matrix elements of $\hat H_{eff.}$. Examples of such diagrams are shown in   Fig.~\ref{fig:diagramV}.   In practice, all these diagrams are generated automatically by the projection code.

 \begin{figure}[]
   \centering
   \subfigure[]{\label{fig:symmetrical_eff_charge_int}\includegraphics[width=0.25\textwidth , angle=270]{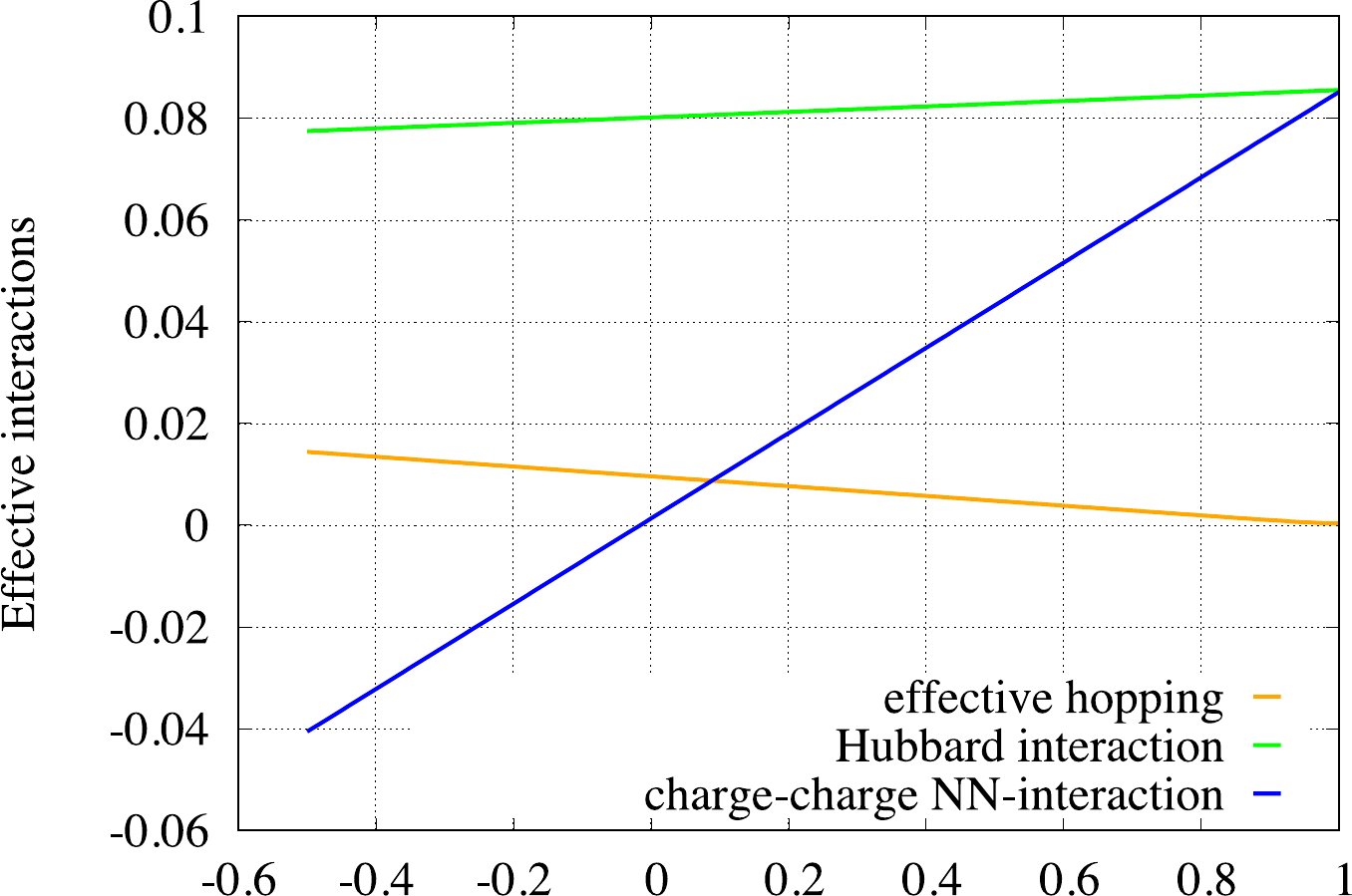}}
   \subfigure[]{\label{fig:symmetrical_spin_charge_int}\includegraphics[width=0.25\textwidth , angle=270]{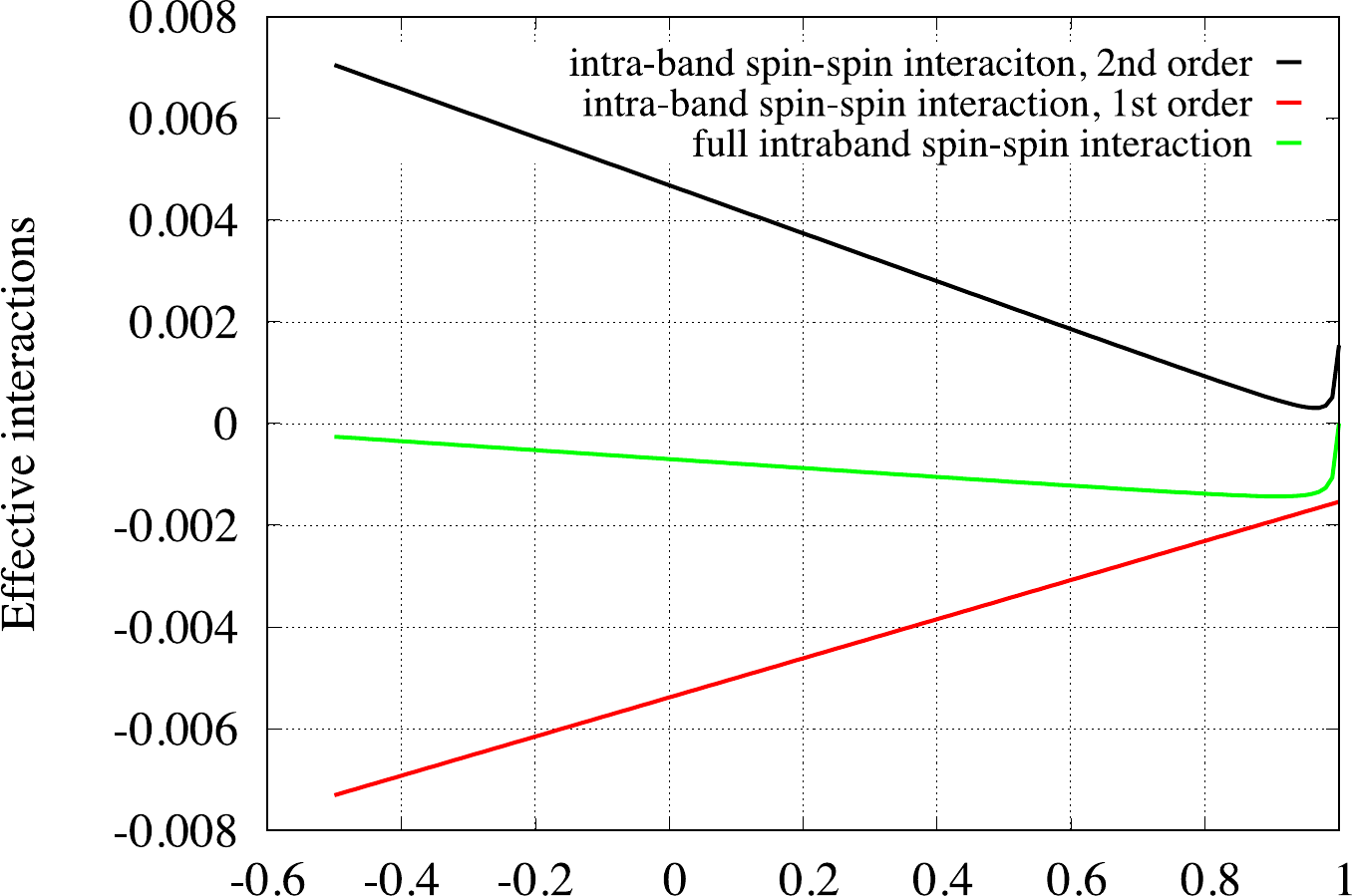}}
        \caption{Effective interactions for two electrons in the flat band, computed from intra-band term $\hat H_\phi$ within the second order perturbation theory. (a) effective hopping $t_{eff.}$ (see \ref{eq:hop_eff}); effective Hubbard interaction $U_{eff.}$ (see \ref{eq:U_eff_def}) and full effective charge-charge interaction $V_{eff.} + \frac{1}{2} J^{(1)}_{eff.}$ (see \ref{eq:V_eff_def}). (b) Effective spin-spin interaction within the 1st ($J^{(1)}_{eff.}$, see \ref{eq:J1_eff_def}) and the 2nd order ($J^{(2)}_{eff.}$, see \ref{eq:J2_eff_def}) perturbation theory, as well as their sum $J^{(2)}_{eff.} + J^{(2)}_{eff.}$. Calculation were done for the system described in the figure \ref{fig:system1_scheme} with the matrix of interactions defined in \ref{eq:U_matrix}. $U_0=1.0$. 
           }
   \label{fig:system1_analytical}
\end{figure}

In order to further simplify the calculations, we take into account only single-particle excitations, described by $\hat V^{(1)}$ operator when writing the nominator of the second term in \ref{eq:eff_ham1}. Many-particle excitations in the intermediate state $| \Psi_I \rangle$ in the second term in \ref{eq:eff_ham1} have larger energy described by the term $\langle \Psi_I | \hat H^{(4)}| \Psi_I \rangle$ again due to the presence of a large mass gap in the system. Thus, these excitations are again suppressed by the mass gap. 

The final expression for the effective Hamiltonian can be written as
\begin{eqnarray}
     \hat H_{eff.}(E)\approx \hat H_\phi + \hat P \hat V^{(1)} \hat R \frac{1}{E-\hat R \hat {\tilde H}_\phi \hat R} \hat R \hat V^{(1)} \hat P,
    \label{eq:eff_ham_fin}
\end{eqnarray}
where 
\begin{eqnarray}
    \hat {\tilde H}_\phi=\hat {\tilde H}^{(1)} + \hat {\tilde H}^{(2)} + \hat {\tilde H}^{(3)} + \hat H^{(4)} + \hat H^{(5)}
    \label{eq:H_phi_tilde}
\end{eqnarray}
is the part of $\hat {\tilde H}_\phi$ operator, diagonal in the basis \ref{eq:wf_general}.

Subsequently, we construct  the  matrix of the Hamiltonian $\hat H_{eff.}$ in the subspace \ref{eq:wf_projected} taking into account all possible single electron (hole) excitations in the intermediate state. After construction of this matrix, the eigenvalues are found using   Eq.~\ref{eq:det_eq}.

 \begin{figure}[]
   \centering
   \subfigure[]{\label{fig:system1_fitting_scheme}\includegraphics[width=0.12\textwidth , angle=90]{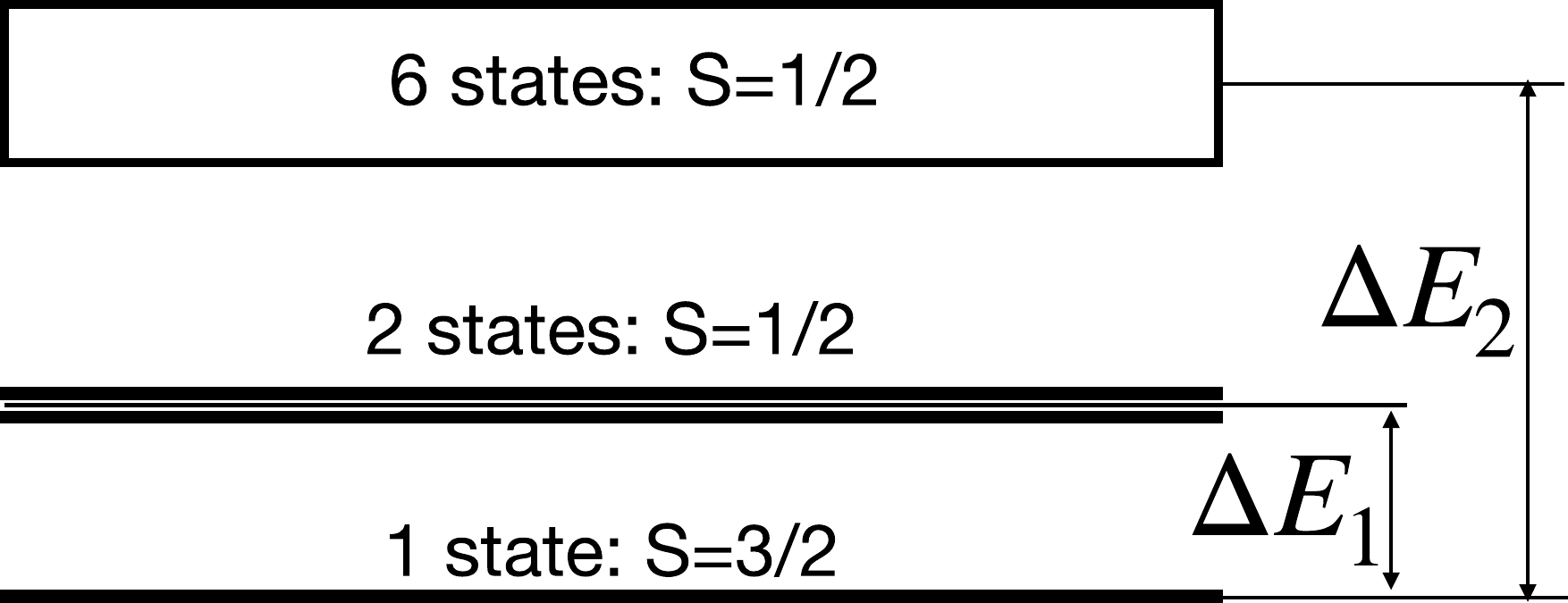}}
   \subfigure[]{ \label{fig:system1_QMC_T_profiles}\includegraphics[width=0.25\textwidth , angle=270]{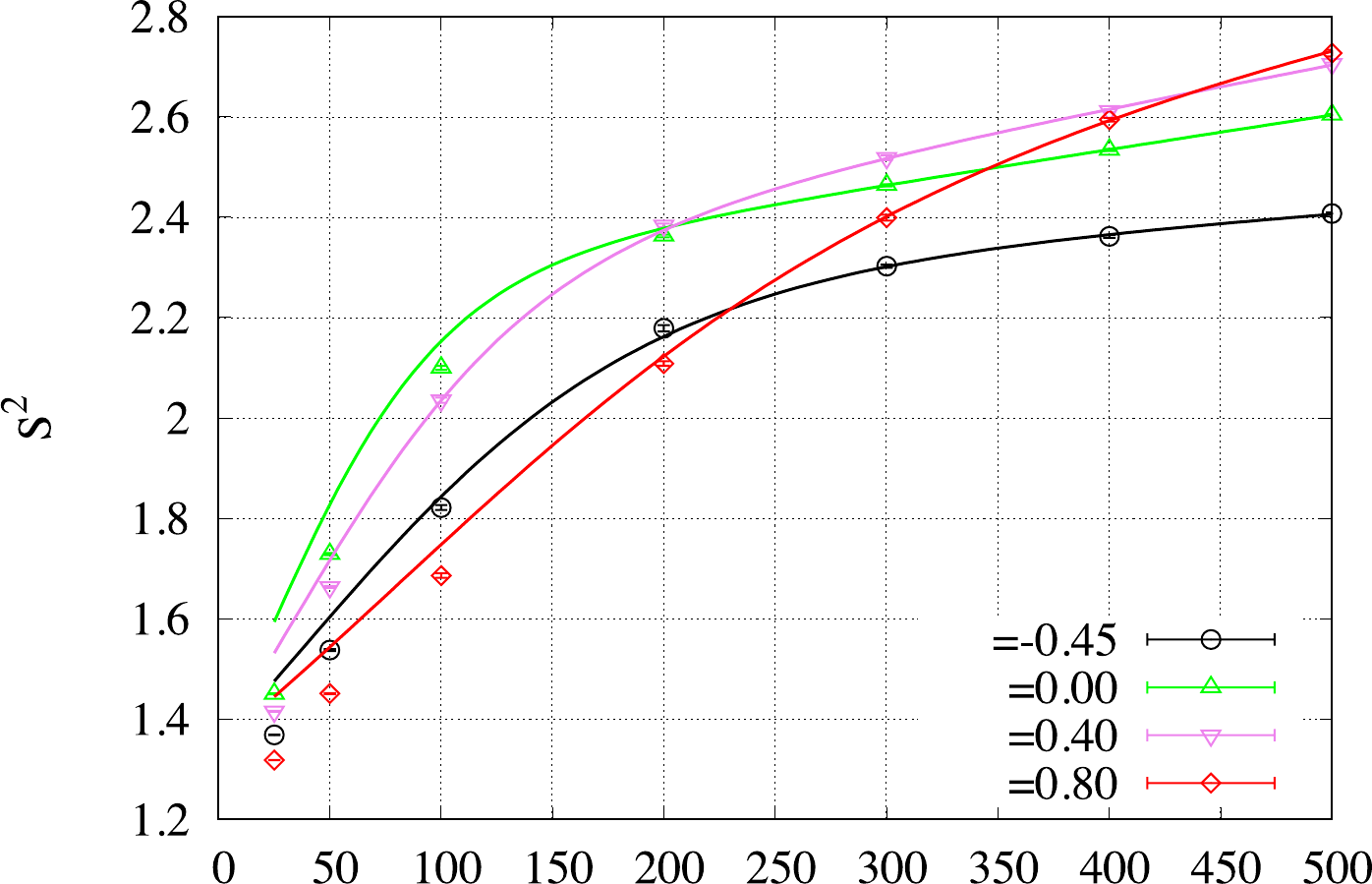}}
         \caption{(a) Scheme of the lowest energy levels of the approximate effective Hamiltonian from eq. \ref{eq:eff_ham_fin} for the system described in the figure \ref{fig:system1_scheme} leading to the $\bar S(\beta)$ function \ref{eq:fitting_curve} used for the fitting in the figure (b). Narrow band is depicted by a rectangle, two degenerate states are depicted as double line. Trivial $M_z$ degeneracy of the spin states is not shown on the scheme. (b)Temperature profiles of $\langle \hat S^2 \rangle$ from QMC data and their fitting with two-exponents function $\bar S(\beta)$ \ref{eq:fitting_curve}. Calculations were done for the system described in the figure \ref{fig:system1_scheme} with the matrix of interactions defined in \ref{eq:U_matrix}. $U_0=0.5$. }
\end{figure}

\subsection{\label{sec:GenProp}Approximate solution for intra-band terms of effective Hamiltonian}

The two terms in  the effective Hamiltonian  of  Eq.\ref{eq:eff_ham_fin} can be considered separately. The first term does not contain any excitations to the upper or lower bands, thus we collectively call all terms in $\hat H_\phi$ as "intra-band". The second term in   \ref{eq:eff_ham_fin} contains excitations to upper and lower band, thus we call all terms within it as "exchange terms".  Before moving to the exact diagonalization of the approximate effective Hamiltonian \ref{eq:eff_ham_fin}, we consider a  perturbative   approach  to provide simplified approximate expressions of the effective interactions between electrons inside the flat band. Namely, we concentrate on $\hat H_\phi$ Hamiltonian ("intra-band" terms) only, and develop the separate perturbation theory for it assuming that the overlap between Wannier functions concentrated near different adatoms is small. 

In order to provide an explicit example, we consider a system of three adatoms depicted in  Fig.~\ref{fig:system1_scheme}.  The interaction Hamiltonian includes three sites  of  the Honeycomb lattice at  which  the   amplitudes of  the  corresponding  Wannier  functions   are   maximal.   With this  arrangement,  the matrix of interactions includes just these three sites and is written as
\begin{eqnarray}
    U= U_0 \begin{pmatrix}
1 & \alpha & \alpha\\
\alpha & 1 & \alpha\\
\alpha & \alpha & 1
\end{pmatrix}.
    \label{eq:U_matrix}
\end{eqnarray}
The parameter $\alpha$ changes the ratio between on-site and long-range interactions. The matrix $U$  is  positive  definite  
provided  that  $\alpha \in [-1/2; 1]$, and only this interval  is  accessible to QMC simulations. The full Hamiltonian $\hat H = \hat H_K + \hat H_U$ thus reads as
\begin{equation}
  \hat{H}  =    \hat{H}_K  +  \frac{1}{2}\sum_{i,j=1}^{3} U_{i,j}   \hat{q}_i  \hat{q}_j 
  \label{eq:H_QMC.eq}
\end{equation}
with $  \hat{q}_i  =  \sum_{\sigma} ( \hat{a}^{\dagger}_{i,\sigma} \hat{a}^{\phantom\dagger}_{i,\sigma} - 1/2)  $.
The position of the  adatoms and the values of hoppings are given in  Fig.~\ref{fig:system1_scheme}. 

 For the model of  Eq.~\ref{eq:H_QMC.eq}  $\hat H_\phi$  is constructed   with  \ref{eq:H_psi_V_def}.  Terms proportional to $H^{(i)}, i\neq 1$ give only two-fermionic terms for the flat band operators $\hat \phi_{i, \sigma}$. These terms  corresponds  to  a chemical potential for the flat band  and 
 ensure   half-filling. Non trivial electron-electron correlations   stem  from the  four-fermionic terms proportional to $H^{(1)}_{ijkn}$.  A hierarchy   in these  terms  emerges  by   assuming that the overlap between different Wannier functions inside the flat band is small. This assumption can be formalized in the following relations between   the coefficients $H^{(1)}_{ijkn}$
\begin{eqnarray}
|H^{(1)}_{iiii}|>|H^{(1)}_{iijj}|\gg|H^{(1)}_{ijji}|, \, |H^{(1)}_{iiij}| \nonumber \\ \gg |H^{(1)}_{ikkj}| \gg |H^{(1)}_{ijkn}|, i\neq j \neq k \neq n
\label{eq:H_psi_pert_th}
\end{eqnarray}

This  implies  that the dominant term in the expression  \ref{eq:H_psi_V_def} for $\hat H_\phi$ is the effective Hubbard interaction  
\begin{eqnarray}
\hat H^{(H)}_{eff.} = \sum_i U_{eff.}  \hat \phi^\dag_{i, \uparrow} \hat \phi_{i, \uparrow}  \hat \phi^\dag_{j, \downarrow} \hat \phi_{j, \downarrow},
    \label{eq:eff_hubbard}
\end{eqnarray}
where 
\begin{eqnarray}
U_{eff.} = H^{(1)}_{iiii}, \quad \forall  i \label{eq:U_eff_def}.
\end{eqnarray}
 In general,  $U_{eff.}$  should be dependent on the index of corresponding Wannier wave function. However, all three adatoms are equivalent in our case, so that $U_{eff.}$ does not depend on the choice of particular state in the flat band. Thus, for simplicity, we drop the $i$ index on the left hand side of the expression \ref{eq:U_eff_def}.

As  mentioned  previously, global   SU(2)  spin    and  U(1)  charge  symmetries  are  conserved  in the projected   Hilbert  space,   spanned  by  the  states  $| \Psi_P \rangle$ (see  Eq.~\ref{eq:wf_projected}).  Hence we can use the third component of total spin $M_z$ and total number of particles to classify the eigenstates. These symmetries together with the dominance of the effective Hubbard interaction \ref{eq:U_eff_def} ensure that the ground state belongs to the subspace corresponding to half filled flat band with three electrons in it. It is also enough to only consider $ M_z= 1/2 $.

 \begin{figure}[t]
   \centering
   \subfigure[]{\label{fig:system1_U0.5_comparison}\includegraphics[width=0.25\textwidth , angle=270]{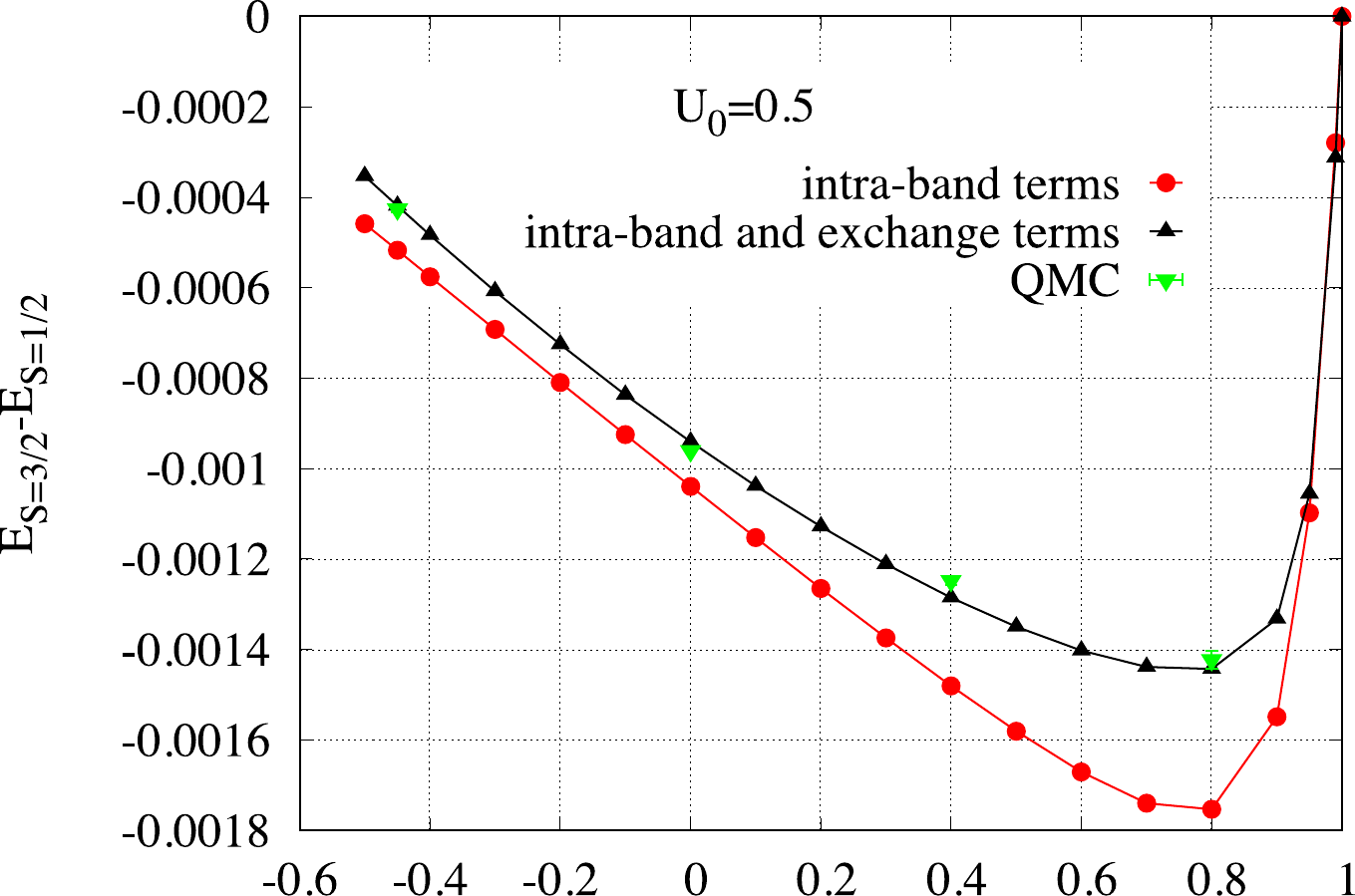}}
   \subfigure[]{\label{fig:system1_U1.0_comparison}\includegraphics[width=0.25\textwidth , angle=270]{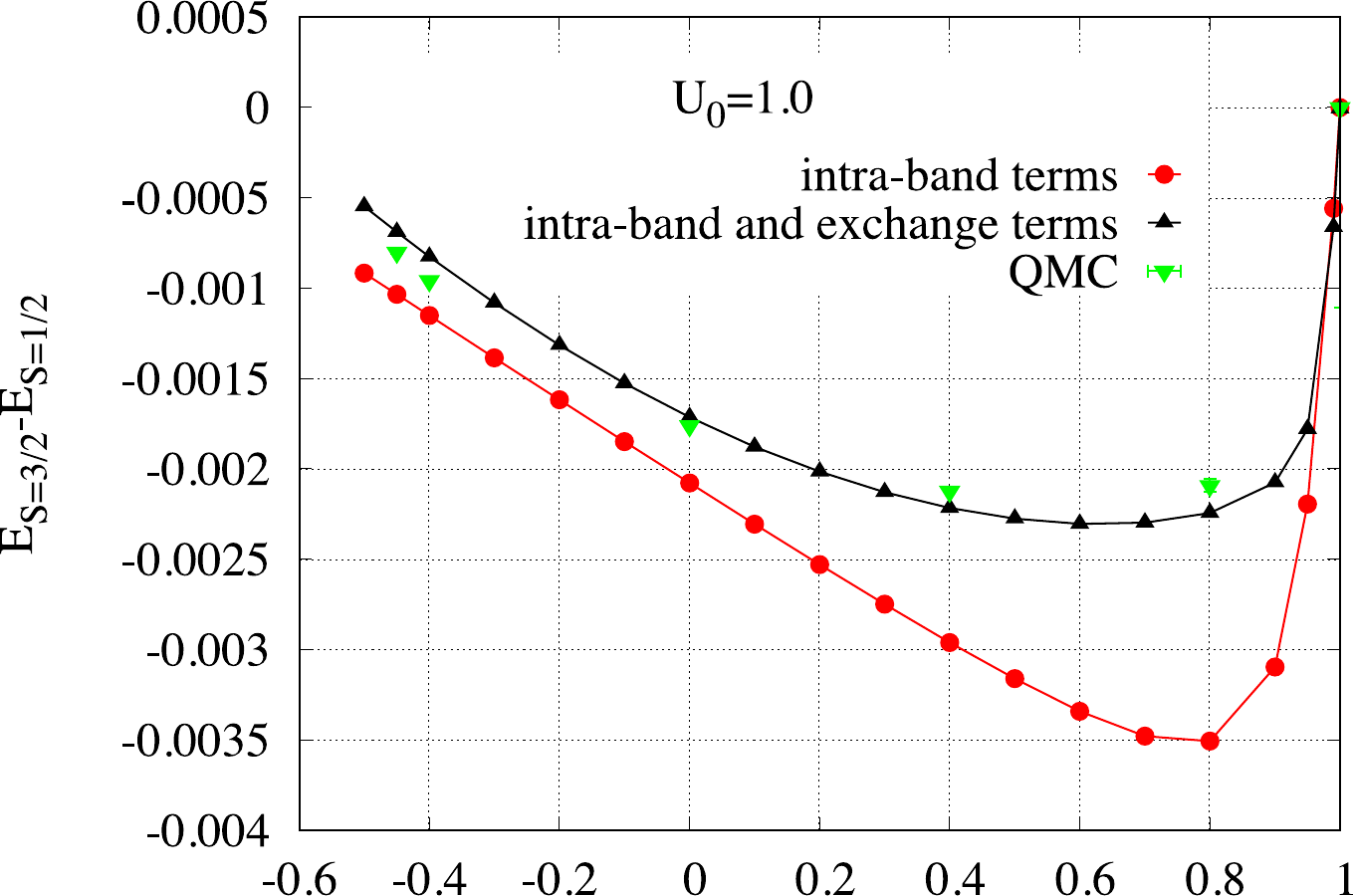}}
      \subfigure[]{\label{fig:system1_U2.0_comparison}\includegraphics[width=0.25\textwidth , angle=270]{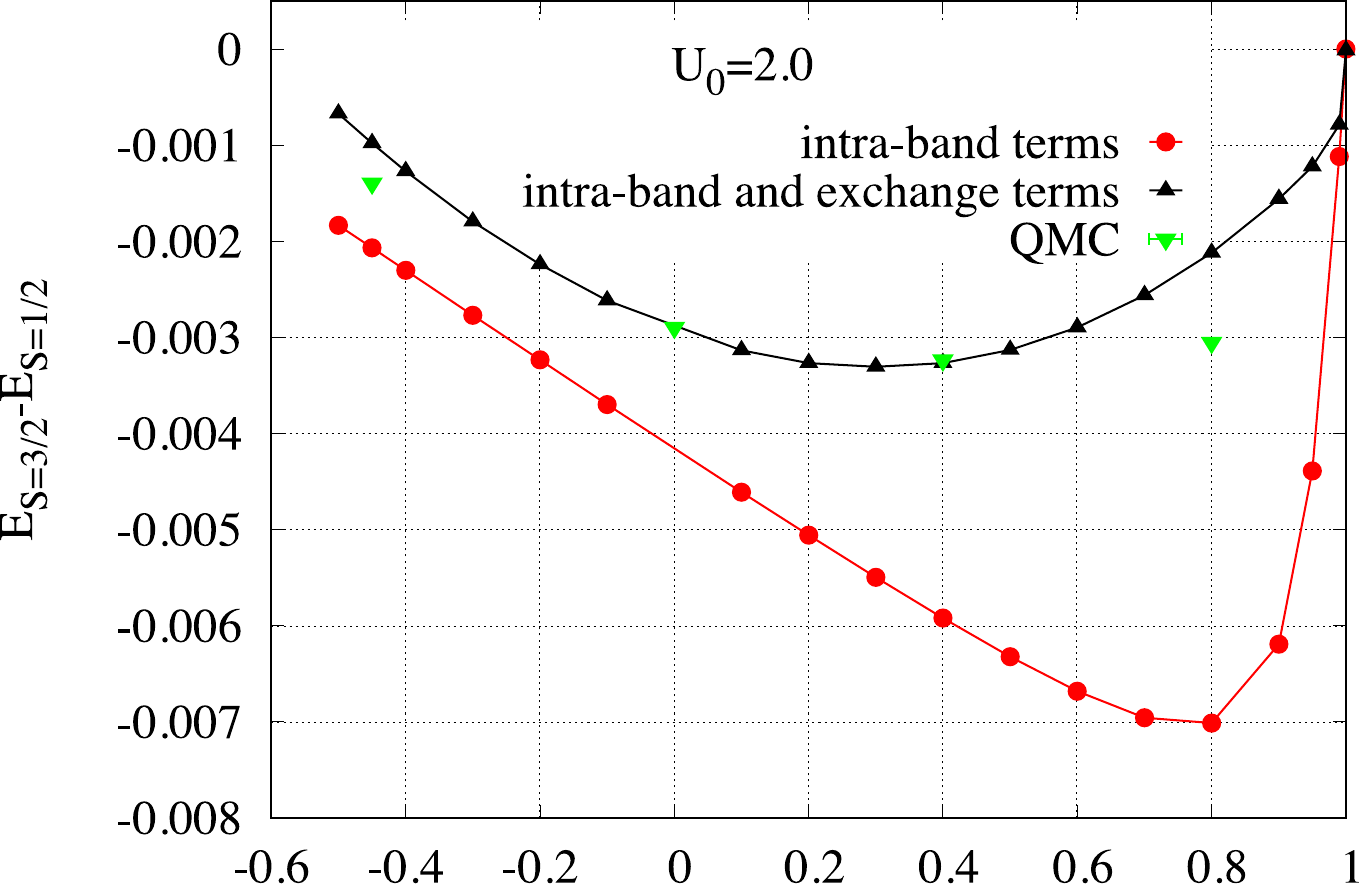}}
       \caption{Energy gap between  $S=3/2$ and $S=1/2$ states in the subspace with three electrons in the flat band for the system shown in the figure \ref{fig:system1_scheme}. Matrix of interactions is defined in \ref{eq:U_matrix}. Intra-band terms correspond to the diagonalization of $\hat H_\phi$ in \ref{eq:eff_ham_fin}.  In the case of exchange terms, only single electron (hole) excitations are taken into account (see the second term in \ref{eq:eff_ham_fin}).}
   \label{fig:system1_comparison}
\end{figure}

The next term in the hierarchy of four-fermionic terms in $\hat H_\phi$ is the effective charge-charge interaction. 
\begin{eqnarray}
    \hat H^{Q}_{eff.}= \frac{1}{2} V_{eff.} \sum_{i \neq j}   \hat Q_i   \hat Q_j,
    \label{eq:eff_Q}
\end{eqnarray}
where 
\begin{eqnarray}
   V_{eff.} =H^{(1)}_{iijj}, \quad \forall  i \neq j.
    \label{eq:V_eff_def}
\end{eqnarray}
$\hat Q_i=\hat \phi^\dag_{i, \uparrow} \hat \phi_{i, \uparrow} +\hat \phi^\dag_{i, \downarrow} \hat \phi_{i, \downarrow}  - 1$ is the charge operator for the electron in the flat band at the i-th state. 

Both $\hat H^{Q}_{eff.}$ and $\hat H^{H}_{eff.}$ are diagonal in the basis  \ref{eq:wf_projected}. Thus, according to the assumption \ref{eq:H_psi_pert_th} we can use them as the zeroth order term in the perturbation theory for $\hat H_\phi$:
\begin{eqnarray}
    \hat H^{(0)}_{eff.}=\frac{1}{2} U_{eff.} \sum_i \hat Q^2_i + \frac{1}{2} V_{eff.} \sum_{i \neq j}   \hat Q_i   \hat Q_j.
    \label{eq:0th_order}
\end{eqnarray}
Since typically $U_{eff.}>V_{eff.}$, the zeroth order of perturbation theory results in fully localized electrons inside the flat band. 

All other terms in $\hat H_\phi$ are considered as a perturbation in comparison to \ref{eq:0th_order}. In the first order perturbation theory, only terms proportional to $H^{(1)}_{ijji}$ contribute and we obtain an effective spin-spin and additional charge-charge interactions between electrons in the flat band:
\begin{eqnarray}
\hat H^{(1)}_{eff.}= J^{(1)}_{eff.} \sum_{i\neq j} \left( 2  \hat{\vec S}_i \hat{\vec S}_j  + \frac{1}{2} \hat Q_i \hat Q_j \right),
    \label{eq:1st_order}
\end{eqnarray}
where 
\begin{eqnarray}
J^{(1)}_{eff.}  = - H^{(1)}_{ijji},  \quad \forall i \neq j
    \label{eq:J1_eff_def}
\end{eqnarray}
At   second order, we obtain additional spin-spin interactions
\begin{eqnarray}
\hat H^{(2)}_{eff.}= J^{(2)}_{eff.} \sum_{i\neq j} \left( 2  \hat{\vec S}_i \hat{\vec S}_j \right),
    \label{eq:2nd_order}
\end{eqnarray}
where
\begin{eqnarray}
J^{(2)}_{eff.}  = \frac{|H^{(1)}_{iiij} -H^{(1)}_{jjji}|^2 + \sum_{k \neq i,j} |H^{(1)}_{ikkj}|^2}{U_{eff.} - (V_{eff.} + \frac{1}{2} J^{(1)}_{eff.})} \quad \forall  i \neq j.
    \label{eq:J2_eff_def}
\end{eqnarray}
This term is   similar to the expression we obtain  for the effective Heisenberg model  in terms of  a   strong-coupling expansion of the Hubbard model. 
Due to this analogy, we can introduce  an effective hopping, induced by interactions inside the flat band:
\begin{eqnarray}
t_{eff.}  = \sqrt{ |H^{(1)}_{iiij} - H^{(1)}_{jjji}|^2 + \sum_{k \neq i,j} |H^{(1)}_{ikkj}|^2 }, \quad \forall  i \neq j.
    \label{eq:hop_eff}
\end{eqnarray}
Dependencies of the effective interactions on $\alpha$ for $U_0=1$ are shown in the Fig.\ref{fig:system1_analytical} for the system described in  Fig.~ \ref{fig:system1_scheme}. We plot the  effective Hubbard interaction $U_{eff.}$, full effective charge-charge interaction $V_{eff.} + \frac{1}{2} J^{(1)}_{eff.}$, and the first and the second order spin-spin interactions $J^{(1)}_{eff.}$ and $J^{(2)}_{eff.}$ separately. 

As one can see, the Hubbard interaction is indeed dominant for most of  the  considered values of  $\alpha$, except for the region around $\alpha=1$, where the nearest-neighbour charge-charge interaction  competes with Hubbard interaction. Thus the perturbation theory for intra-band terms is justified everywhere except the vicinity of the $\alpha=1$ point. 
Comparison of $J^{(1)}_{eff.}$ and $J^{(2)}_{eff.}$ shows that non-frustrated negative $J^{(1)}_{eff.}$ interaction dominates  over frustrated positive $J^{(2)}_{eff.}$ interaction everywhere within the region accessible  to QMC   simulations:  $\alpha \in [-1/2; 1]$. Of course, this is only a very preliminary result, which only takes into account intra-band terms, and relies on perturbation theory. However, this picture will be justified later for the complete exact diagonalization of  the approximate  effective Hamiltonian \ref{eq:eff_ham_fin} including exchange terms.

  \begin{figure}[]
   \centering
 \includegraphics[width=0.3\textwidth , angle=0]{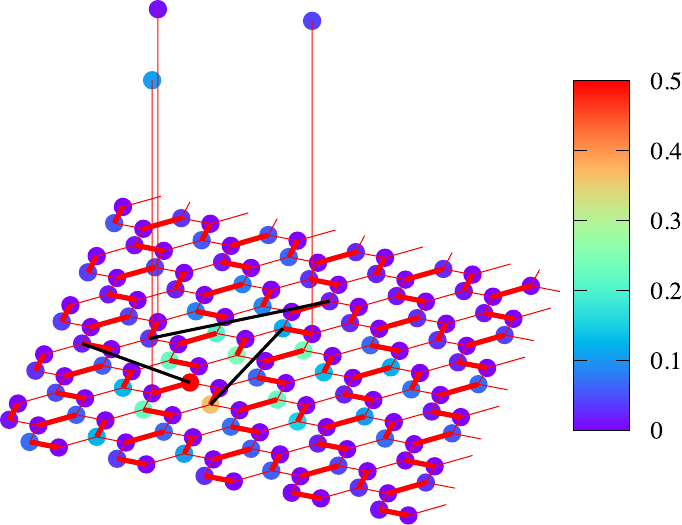}
             \caption{Scheme of the lattice model with three additional sites ("adatoms") and alternative setup for interaction Hamiltonian  \ref{eq:int_hamiltonian}. Red lines represent hoppings, Kekule pattern is shown with bold and thin lines.  Bold ones  correspond to larger hopping $t=1.3$ and the thin lines correspond to smaller hopping $t=1.0$. Hoppings to additional sites are equal to $t'=10.0$.  Black lines connect the sites involved in charge-charge interactions. Color scale corresponds to the modulus of the Wannier wavefunction, concentrated around one of the additional sites.    }
   \label{fig:system2_scheme}
\end{figure}

 \begin{figure}[]
   \centering
   \subfigure[]{\label{fig:nonsymmetrical_eff_charge_int}\includegraphics[width=0.25\textwidth , angle=270]{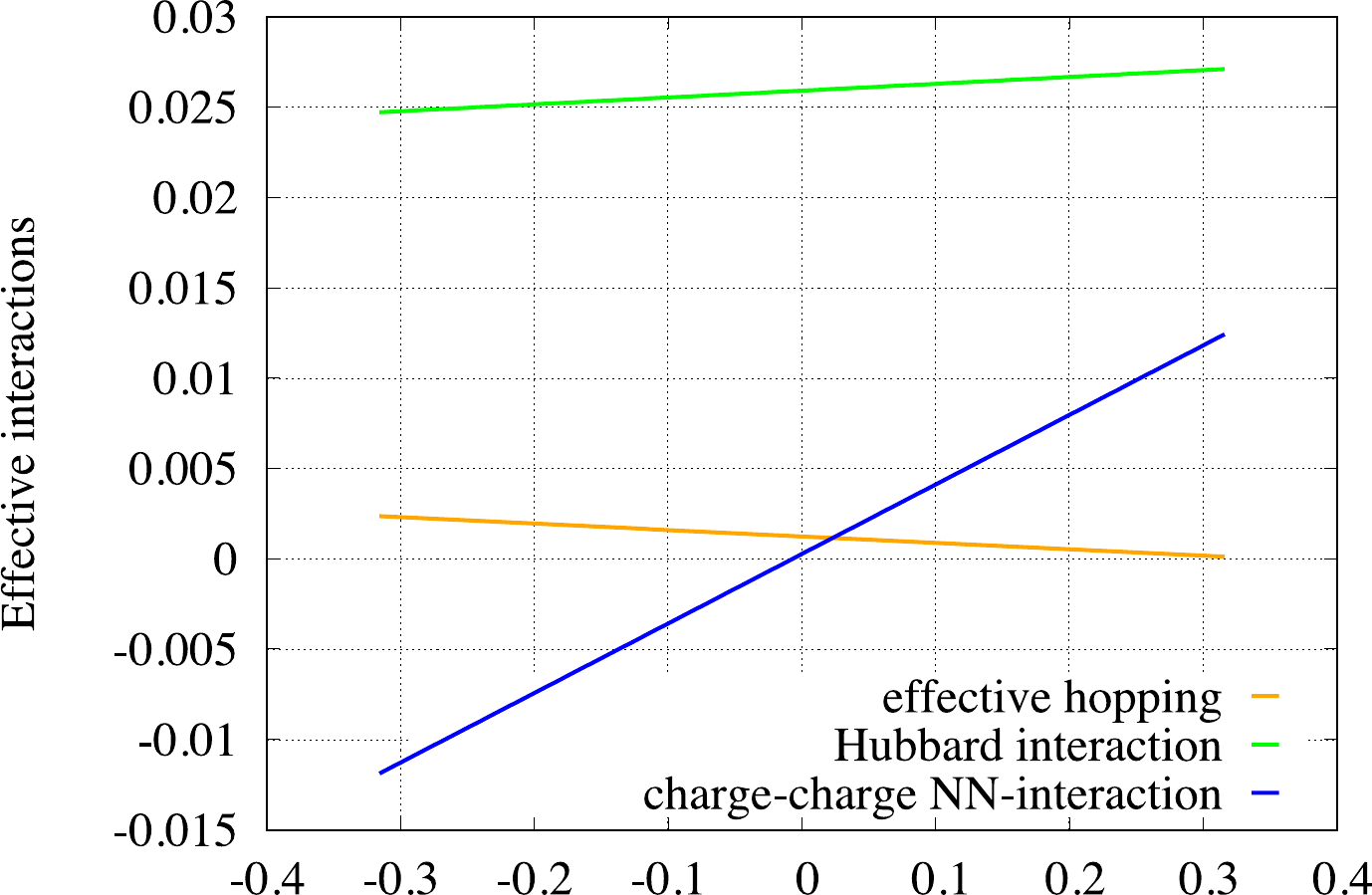}}
   \subfigure[]{\label{fig:nonsymmetrical_spin_charge_int}\includegraphics[width=0.25\textwidth , angle=270]{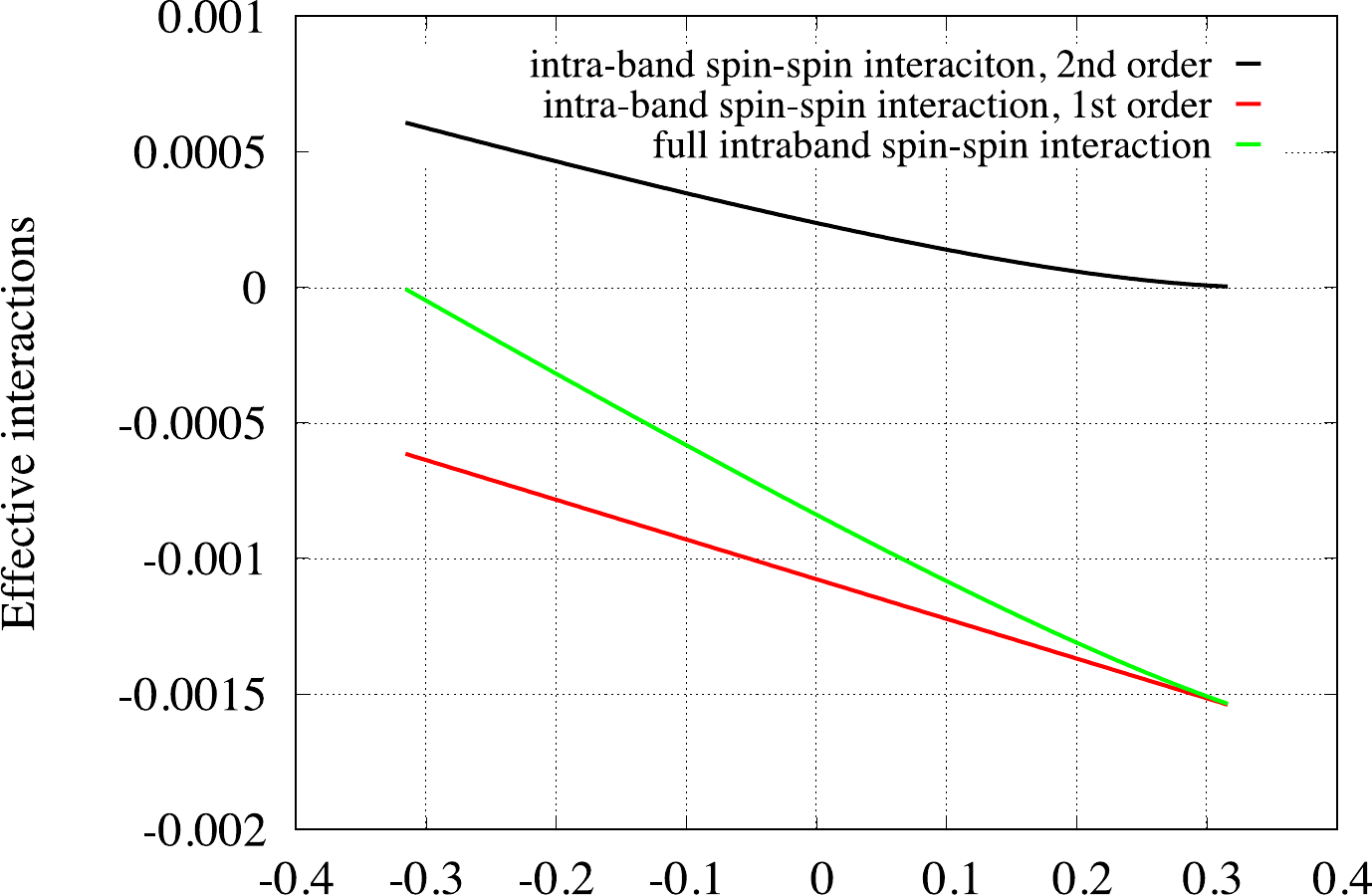}}
        \caption{Effective interactions for two electrons in the flat band, computed from intra-band term $\hat H_\phi$ within the second order perturbation theory for the system described in the figure \ref{fig:system2_scheme}.  (a) effective hopping $t_{eff.}$ (see \ref{eq:hop_eff}); effective Hubbard interaction $U_{eff.}$ (see \ref{eq:U_eff_def}) and full effective charge-charge interaction $V_{eff.} + \frac{1}{2} J^{(1)}_{eff.}$ (see \ref{eq:V_eff_def}). (b) Effective spin-spin interaction within the 1st ($J^{(1)}_{eff.}$, see \ref{eq:J1_eff_def}) and the 2nd order ($J^{(2)}_{eff.}$, see \ref{eq:J2_eff_def}) perturbation theory, as well as their sum $J^{(2)}_{eff.} + J^{(2)}_{eff.}$. Calculation were done with the matrix of interactions defined in \ref{eq:u_m_system2}. $U_0=1.0$. 
       }
   \label{fig:system2_analytical}
   
\end{figure}

\section{\label{sec:ALFVer} QMC verification of the projection algorithm}

In order to estimate the quality of the approximations made during the derivation of the effective Hamiltonian \ref{eq:eff_ham_fin} a verification is made via comparison to  unbiased QMC simulations   carried  out  for parameters of the model, where the sign problem is absent. For this verification, we switch to the exact diagonalization of the approximate effective Hamiltonian \ref{eq:eff_ham_fin}, including both intra-band and exchange terms.  

Auxiliary-field QMC simulations were   carried out   with the ALF \cite{ALF2.0}   implementation  of  the  finite  temperature  auxiliary  field  QMC  algorithm \cite{Blankenbecler81,White89,Assaad08_rev}.  We  consider  the system depicted in   Fig.~\ref{fig:system1_scheme} with electron-electron interaction defined in \ref{eq:U_matrix}  and  Hamiltonian   of  Eq.~\ref{eq:H_QMC.eq}. In order to decouple the charge-charge interactions we first rewrite them as the perfect square of one body terms. Depending on the sign of the parameter $\alpha$ we employ different decouplings to prevent the occurrence of the sign problem. For $\alpha$ in the interval $[-1/2; 0]$ we write
 \begin{eqnarray}
\hat H_U=  U_0 \left(\frac{1}{2}+\alpha \right) \sum_{i=1}^{3} \hat{q}_i^2 - \frac{\alpha}{2} U_0 \sum_{i>j} (\hat{q}_i - \hat{q}_j)^2
 \end{eqnarray}
 and for $\alpha \in ]0; 1]$
  \begin{eqnarray}
\hat H_U= \frac{1}{2} U_0 (1-\alpha) \sum_{i=1}^3 \hat{q}_i^2 + \frac{\alpha}{2} U_0 \left( \sum_{i=1}^3 \hat{q}_i \right)^2.
 \end{eqnarray}
Here the sum over $i$ and $j$ goes over the three interacting sites depicted in   Fig.~\ref{fig:system1_scheme}.  After this we can perform a Trotter decomposition and a discrete Hubbard-Stratonovich (HS) decomposition as outlined in Ref.~\cite{ALF2.0}. The Trotter step size was fixed to $\Delta\tau=0.1t$ for all simulations. Lattice size corresponds exactly to Fig.
~\ref{fig:system1_scheme}, with periodical boundary conditions in all directions. 

 Since the (non-)appearance of the frustrated ground state is the most interesting physical effect in this system, we compute the energy difference between the states with $S=3/2$ and $S=1/2$ inside the subspace with 3 electrons in the flat band. This energy difference roughly corresponds to the effective spin-spin interaction between electrons localized around different adatoms, and we use it to compare   the  QMC data with the exact diagonalization of approximate effective Hamiltonian.  

In order to extract this difference from QMC data, we plot the temperature dependence of average squared spin $\bar S = \langle \hat \vec{S}^2 \rangle$ and fit it 
to the  form:
\begin{eqnarray}
    \bar S(\beta) = \frac{9 e^{-\Delta E_2 \beta} + 3 e^{-\Delta E_1 \beta} + 15}{12 e^{-\Delta E_2 \beta} + 4 e^{-\Delta E_1 \beta} +4 }, 
        \label{eq:fitting_curve}
\end{eqnarray}
 where $\beta$ is the inverse temperature.   
Here  $ \vec{S}  =  \frac{1}{2}\sum_{x \in \mathcal{H, A} ,s,s'} a^{\dagger}_{x,s} \vec{\sigma}_{s,s'} a^{\phantom\dagger}_{x,s'}  $    corresponds  to  the total  spin. 
 
 This fitting  form takes into account all states within the subspace with three electrons in the flat band, which is the lowest subspace in the energy spectrum of the whole system. Namely, the fitting  form  accounts  for  one (four-fold degenerate) $S=3/2$ state with energy $E_{S=3/2}$, two (two-fold degenerate) $S=1/2$ states with identical energy $E_{S=1/2}$ and six remaining delocalized two-fold degenerate states with $S=1/2$. The latter states are not completely degenerate and form a narrow band, but for simplicity  we assume that they are characterized by the same energy $E'_{S=1/2}$. Thus the parameters of the fitting   form 
 \ref{eq:fitting_curve} are defined as 
 \begin{eqnarray}
    \Delta E_1 =E_{S=1/2} - E_{S=3/2}, \\
    \Delta E_2 =E'_{S=1/2} - E_{S=3/2}.
     \label{eq:dE_def}
 \end{eqnarray}
The  energy levels are  shown in Fig.~\ref{fig:system1_fitting_scheme}.
Examples of such fits for different values of $\alpha$ are shown in Fig.~\ref{fig:system1_QMC_T_profiles}. 
There are some deviations at low $\beta$, which correspond to the participation of higher energy levels in the $\bar S(\beta)$ profiles. However, at $\beta>200$ the fitting curves are indistinguishable from the QMC data. 

The energy differences $\Delta E_1$ extracted from the fits are compared with the same differences obtained from exact diagonalization of the approximate effective Hamiltonian \ref{eq:eff_ham_fin}. For the latter dataset, we show two points for each $\alpha$: one point corresponds to the exact diagonalization of intra-band terms $\hat H_\phi$ only (an exact version of the approximate calculation from  Sec.~\ref{sec:GenProp}); and another point corresponds to complete Hamiltonian $\hat H_{eff.}$. Difference between these two points show the role of exchange terms. Results are shown in  Fig.\ref{fig:system1_comparison} for three values of $U_0=0.5, 1.0, 2.0$. As expected from the approximations based on the assumption of a large gap, the effective Hamiltonian performs well for smaller $U=0.5, 1.0$, but some noticeable deviations appear at $U=2.0$. However, the approximation never causes a qualitative error: QMC shows that everywhere inside the interval $\alpha \in [-1/2;1]$ accessible for the simulations we get non-frustrated $S=3/2$ ground state, and the same is true for the exact diagonalization of approximate effective Hamiltonian. In general, the comparison shows that it is safe to use the approximate Hamiltonian up to $U/m \simeq 3$, where $m$ is the gap between upper and lower bands.

 \begin{figure}[]
   \centering
   \subfigure[]{\label{fig:system2_fitting_scheme}\includegraphics[width=0.25\textwidth , angle=0]{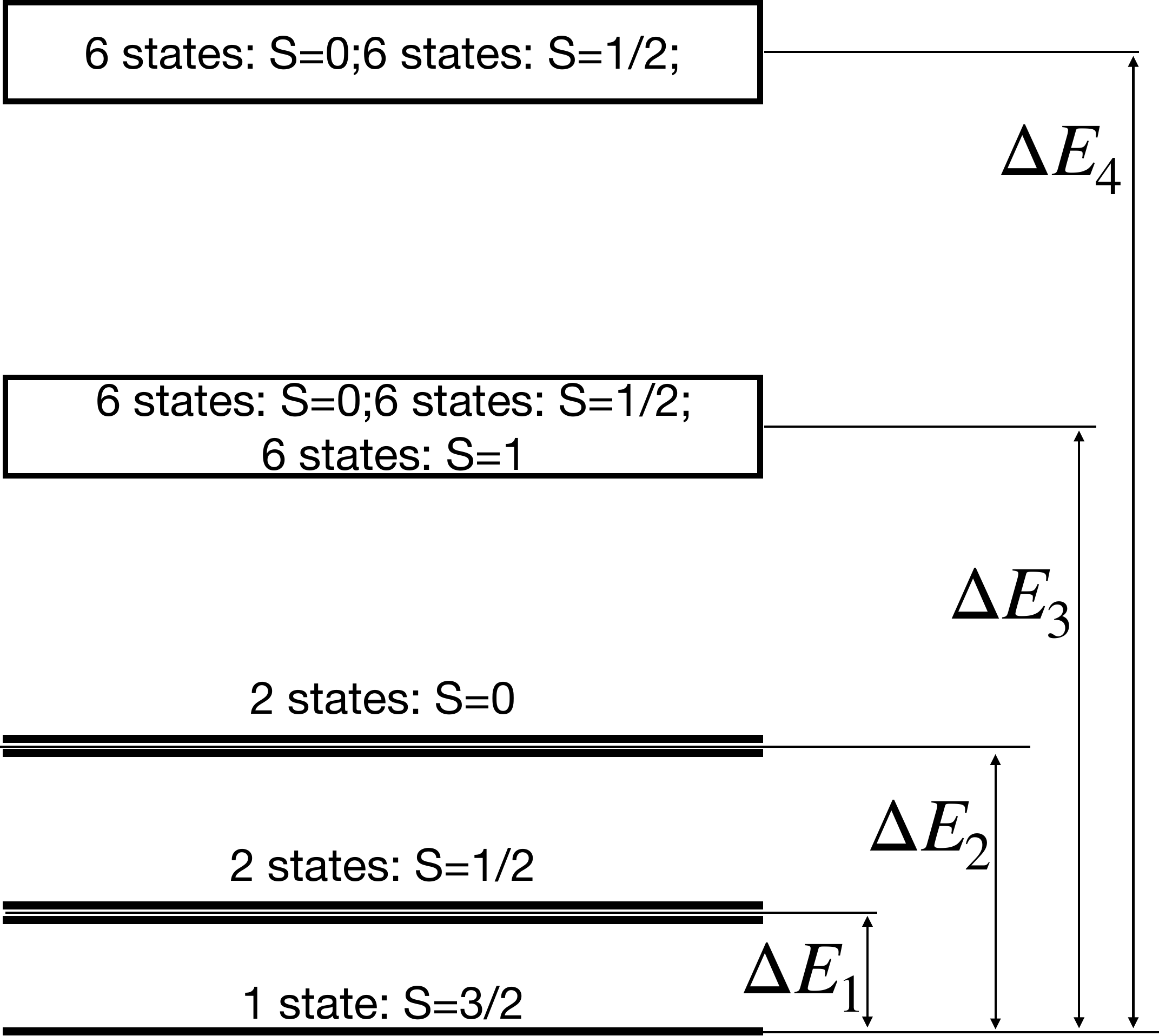}}
   \subfigure[]{\label{fig:system2_QMC_T_profiles}\includegraphics[width=0.25\textwidth , angle=270]{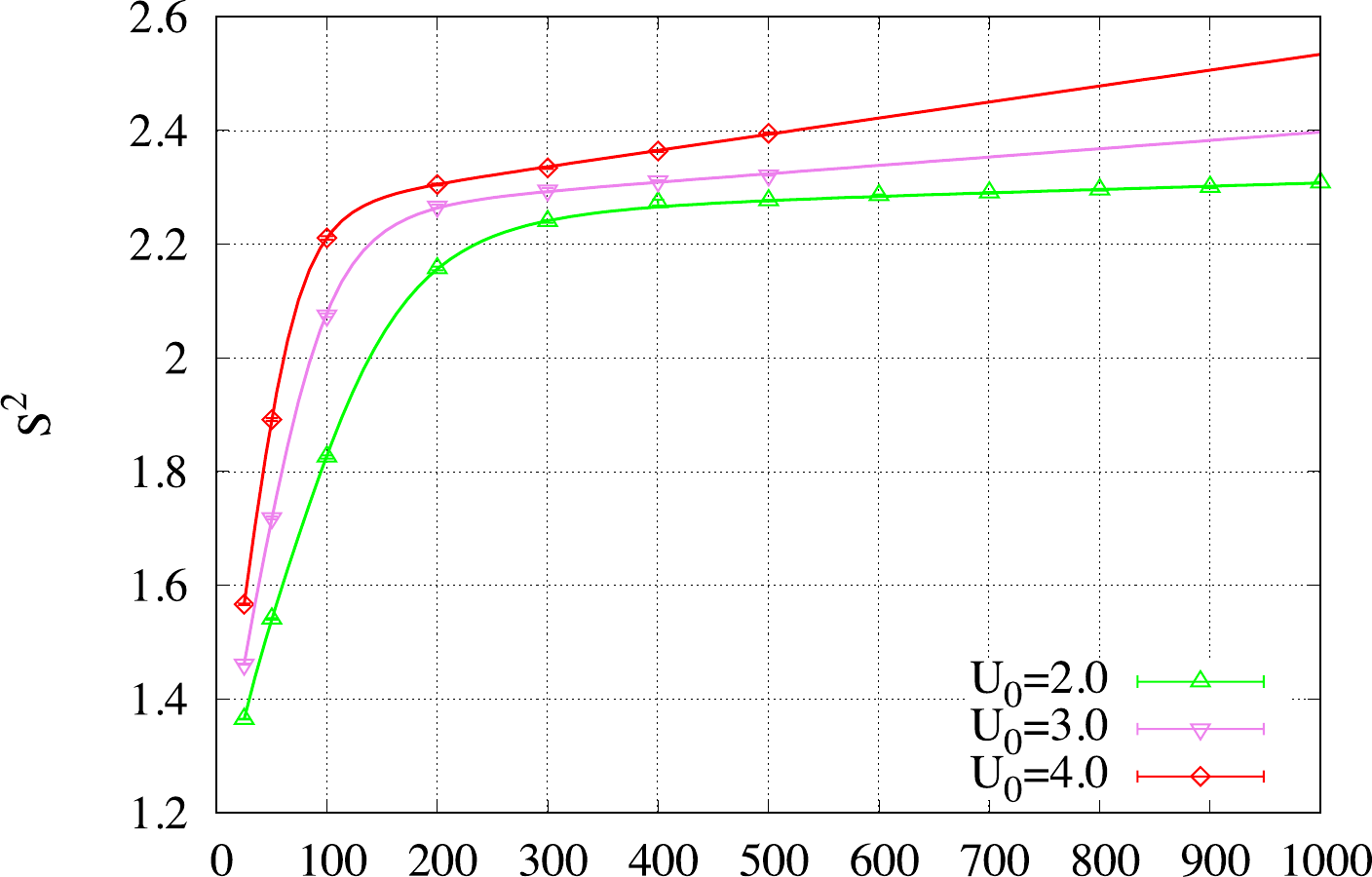}}
         \caption{(a) Scheme of the lowest energy levels of the approximate effective Hamiltonian from eq. \ref{eq:eff_ham_fin} for the system described in the figure \ref{fig:system2_scheme} leading to the $\bar S(\beta)$ function \ref{eq:fitting_curve2} used for the fitting in the figure (b). Narrow bands are depicted by rectangles, two degenerate states are depicted as double line. Trivial $M_z$ degeneracy of the spin states is not shown on the scheme. (b)Temperature profiles of $\langle \hat S^2 \rangle$ from QMC data and their fitting with the four-exponents function $\bar S(\beta)$ \ref{eq:fitting_curve2}. Calculations were done for the system described in the figure \ref{fig:system2_scheme} for the interaction Hamiltonian \ref{eq:interaction_system2} at $\alpha=-\sqrt{1/10}$. }
\end{figure}

  \begin{figure}[]
   \centering
 \includegraphics[width=0.25\textwidth , angle=270]{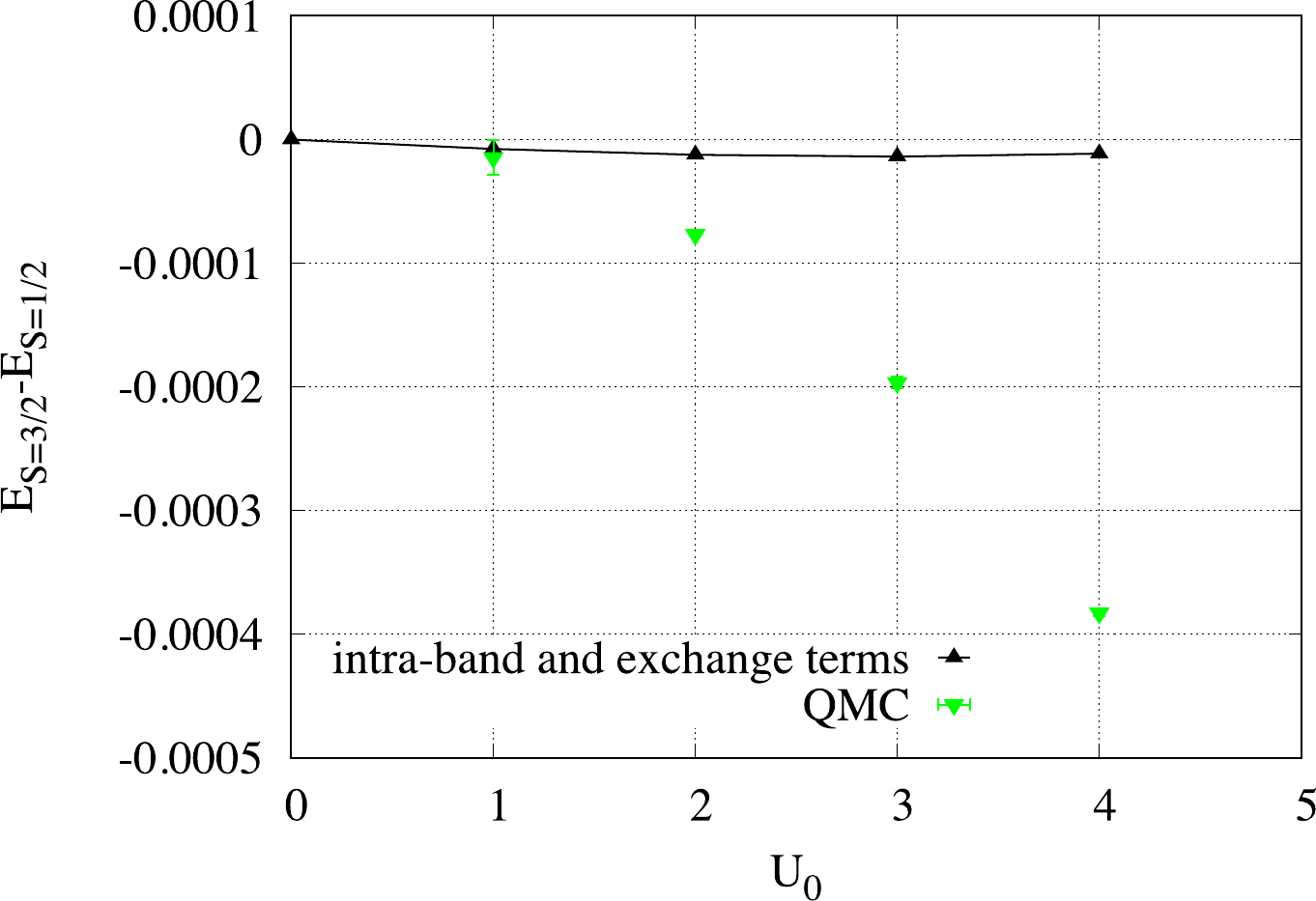}
         \caption{Energy gap between  $S=3/2$ and $S=1/2$ states within the subspace with three electrons in the flat band. Calculations were done for the system shown in the figure \ref{fig:system2_scheme} with the interaction matrix defined in \ref{eq:u_m_system2} at $\alpha=-\sqrt{1/10}$. Intra-band and exchange terms correspond to the diagonalization of the approximate effective Hamiltonian \ref{eq:eff_ham_fin}. QMC data is the result of the fitting procedure, shown in the figure \ref{fig:system2_QMC_T_profiles}.}
   \label{fig:system2_comparison}
\end{figure}

\section{\label{sec:mapping}Mapping of the sign problem for spin frustrated system}

Now the question is, whether we can generate an effective model with frustrated spin interactions  within  the flat band, while still preserving the possibility to perform QMC simulations of the whole lattice model. Unfortunately, attempts of numerical optimization of the lattice model show that the answer is negative.  However, it is possible to generate an effective model with frustrated spins if we consider the microscopic interaction Hamiltonian \ref{eq:int_hamiltonian} with mixture of positive and negative eigenvalues in the matrix $U$. Such models provide a mapping from the sign problem for spin frustrated system to the problem of different signs of eigenvalues of interaction matrix. 

\subsection{\label{sec:mapping_zero}Zero spin-spin interaction at the edge of the parameters' region accessible for QMC}

  \begin{figure}[]
   \centering
 \includegraphics[width=0.3\textwidth , angle=270]{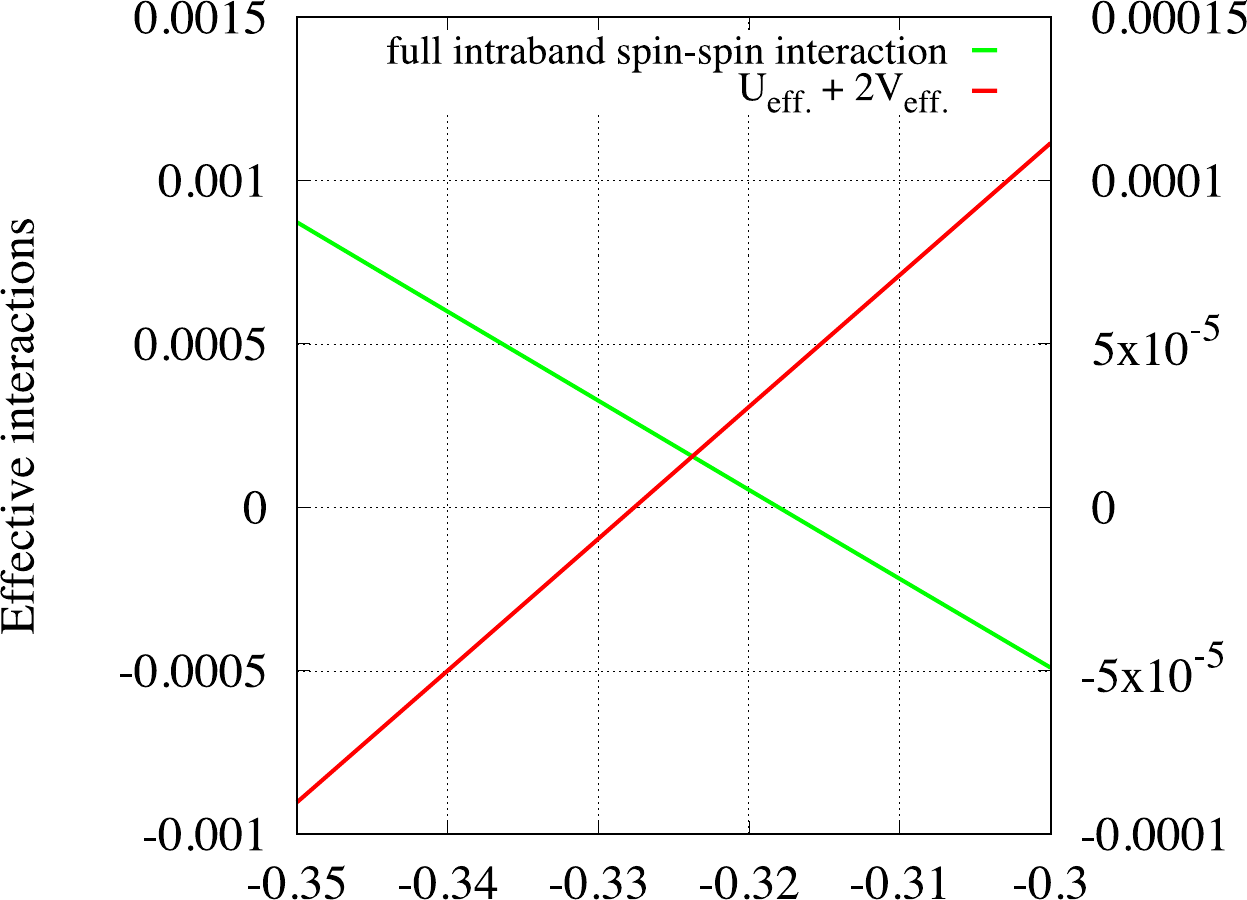}
         \caption{Flat band spin-spin interaction and $U_{eff.} + 2 V_{eff.}$ interaction for the system depicted in the figure \ref{fig:system2_scheme} beyond the critical $\alpha=-\sqrt{1/10}$ value. Left scale corresponds to $U_{eff.} + 2 V_{eff.}$ interaction, right scale corresponds to the spin-spin interaction. Calculations were done for intra-band terms in \ref{eq:eff_ham_fin}, using the second order perturbation theory (see eq. \ref{eq:J1_eff_def} and \ref{eq:J2_eff_def} for spin-spin interaction and eq. \ref{eq:U_eff_def}, \ref{eq:V_eff_def} for charge-charge interaction). Matrix of interactions is defined in \ref{eq:interaction_system2},  \ref{eq:u_m_system2}; $U_0$=1.0.}
   \label{fig:system2_beyond_limit}
\end{figure}

First, we perform a numerical optimization of the $U$ matrix trying to maximize full intra-band spin-spin interaction $J^{(\Sigma)}_{eff.} = J^{(1)}_{eff.} + J^{(2)}_{eff.}$. This quantity is rather cheap to compute, thus the optimization can be performed efficiently. However, it seems that we can not make $J^{(\Sigma)}_{eff.}$ positive keeping all eigenvalues of $U$ matrix also positive.  The maximum which we can obtain is the situation when one (or several) eigenvalues of the U matrix are zero, others are positive and the full intra-band spin-spin interaction is zero. Such  a system is 
shown in Fig.~\ref{fig:system2_scheme}. The interaction consists of three dimer terms, and six different sites are included. These sites are shown in the Fig.~\ref{fig:system2_scheme} connected by black lines. We denote charge operators at these sites as $\hat q^j_i$, where $i=1,2,3$ is index of a dimer and $j=1,2$ is index of a site in the dimer $i$.  $\hat q^1_i$ always corresponds to the site, which is the nearest neighbour to the one with an adatom attached to it (see Fig.~\ref{fig:system2_scheme}). The whole interaction Hamiltonian is written as
\begin{eqnarray}
\hat H_U= \sum^3_{i=1} \frac{1}{2} \sum^2_{n,m=1} U_{nm} \hat q^n_i \hat q^m_i,
\label{eq:interaction_system2}    
\end{eqnarray}
where the $2 \times 2$ matrix $U$ is defined as
\begin{eqnarray}
    U= U_0 \begin{pmatrix}
0.1 & \alpha \\
\alpha & 1  
\end{pmatrix}
  \label{eq:u_m_system2}
\end{eqnarray}
Both eigenvalues of $U$ matrix are positive if $\alpha \in [-\sqrt{1/10};\sqrt{1/10}]$.

Full $\alpha$ profiles of effective interactions obtained via the 2-nd order perturbation theory for intra-band part of the effective Hamiltonian are shown in the Fig.~\ref{fig:system2_analytical}. As one can see, $J^{(\Sigma)}_{eff.}$ is exactly zero at the left edge of the  allowed interval for $\alpha$.

In a second step, we check if the inter-band terms renormalize the effective spin-spin interaction towards frustrated or non-frustrated regime. In addition to that, we again perform QMC simulations at  $\alpha = -\sqrt{1/10}$ in order to provide an independent control to the results of the exact diagonalization of the approximate effective Hamiltonian 
\footnote{In order to decouple the interaction with a HS decomposition for the QMC simulations we rewrote the interaction as 
\begin{eqnarray*}
\hat{H}_U = \frac{1}{2}  U_0 \sum_{i=1}^3 \left(\alpha \hat{q}_i^1+ \hat{q}_i^2 \right)^2
\end{eqnarray*}
 and again fixed $\Delta\tau=0.1t$.
}. 
At this stage we use the routine already described in Sec.~\ref{sec:ALFVer}. The only difference is that due to $S=3/2$ state and $S=1/2$ state being almost degenerate, we need to add another exponent in the fitting form \ref{eq:fitting_curve} and change our assumptions about the excited levels above the ground state. The new fitting function is written as
\begin{widetext}
\begin{eqnarray}
    \bar S(\beta) = \frac{9 e^{-\Delta E_4 \beta} + 45 e^{-\Delta E_3 \beta} + 3 e^{-\Delta E_1 \beta} + 15}{18 e^{-\Delta E_4 \beta} +  36 e^{-\Delta E_3 \beta} + 2 e^{-\Delta E_2 \beta} +4 e^{-\Delta E_1 \beta} +4 }.
    \label{eq:fitting_curve2}
\end{eqnarray}
\end{widetext}
The meaning of the $\Delta E_i$ parameters is shown in the scheme for the energy levels  of  Fig.\ref{fig:system2_fitting_scheme}. This scheme is taken from the diagonalization of the effective Hamiltonian \ref{eq:eff_ham_fin}. The main difference from the previous case \ref{eq:fitting_curve} is that the second excitation after the lowest pair of states ($S=3/2$ and $S=1/2$ with three electrons in the flat band) belongs to the subspace with different particle number. This subspace corresponds to either empty or completely filled flat band with $S=0$. As in the previous case, we are mostly interested again in the gap between $S=3/2$ and $S=1/2$ states. 

Fitting of the QMC data is shown in  Fig.~\ref{fig:system2_QMC_T_profiles}. Since more energy levels are included, the fitting curve describes QMC data even for thew smallest values of $\beta$. Corresponding gaps between $S=3/2$ and $S=1/2$ states extracted from QMC profiles and from the effective Hamiltonian are shown in the   Fig.~\ref{fig:system2_comparison}. As for the first system (see figure \ref{fig:system1_comparison}), QMC is in good agreement with the effective Hamiltonian  up to $U_0=1$ and deviations become noticeable starting from $U_0=2$. 
The main physical result, which we can extract from this plot, is that the exchange terms in the effective Hamiltonian add only small non-frustrated spin-spin interaction in this case, when $\alpha = -\sqrt{1/10}$ and intra-band spin-spin interaction is zero.
Thus, we are unable to   generate a frustrated state without going beyond limiting value of $\alpha$ and introduction of 
negative eigenvalues in the interaction matrix $U$.

 \begin{figure}[]
   \centering
   \subfigure[]{\label{fig:system3_intra-band}\includegraphics[width=0.3\textwidth , angle=270]{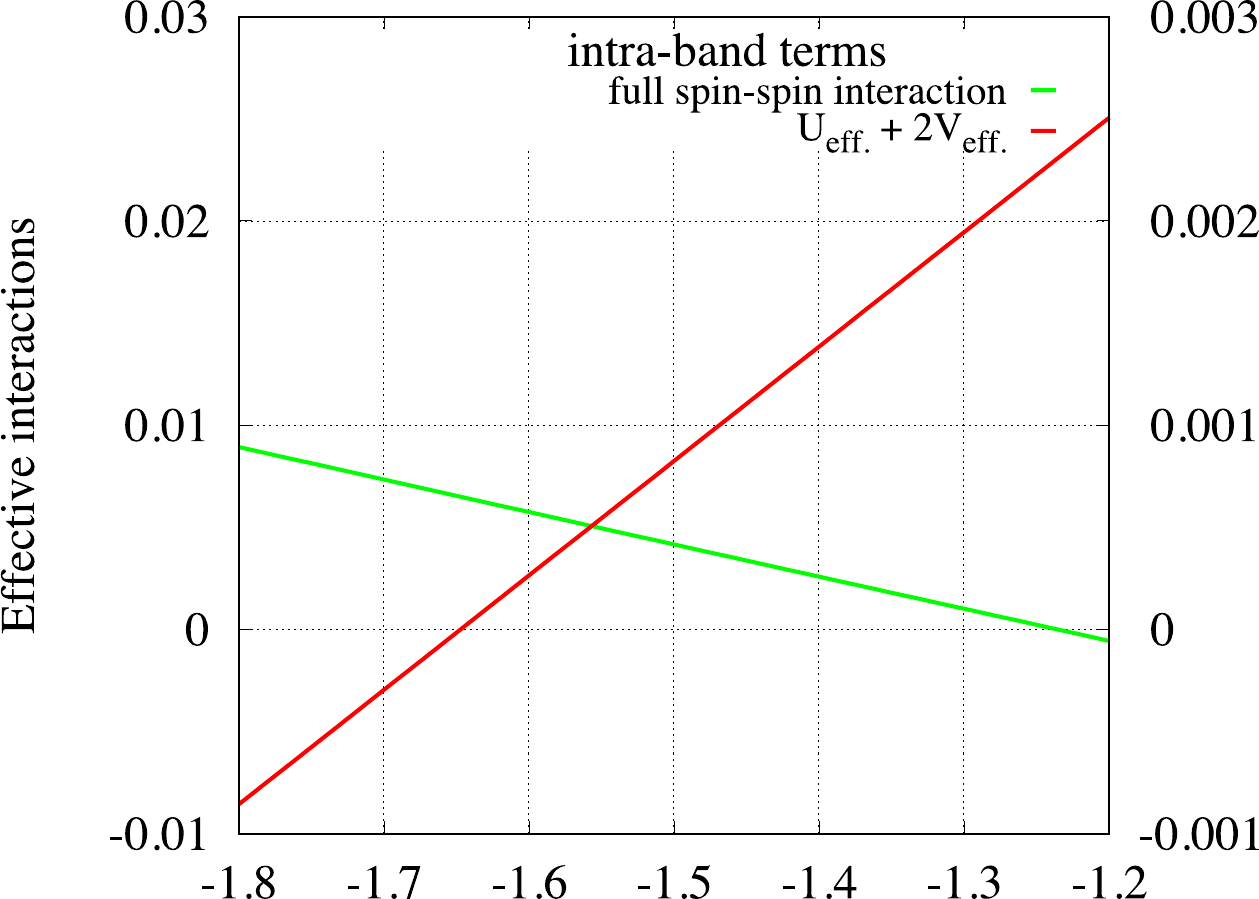}}
   \subfigure[]{\label{fig:system3_inter-band}\includegraphics[width=0.3\textwidth , angle=270]{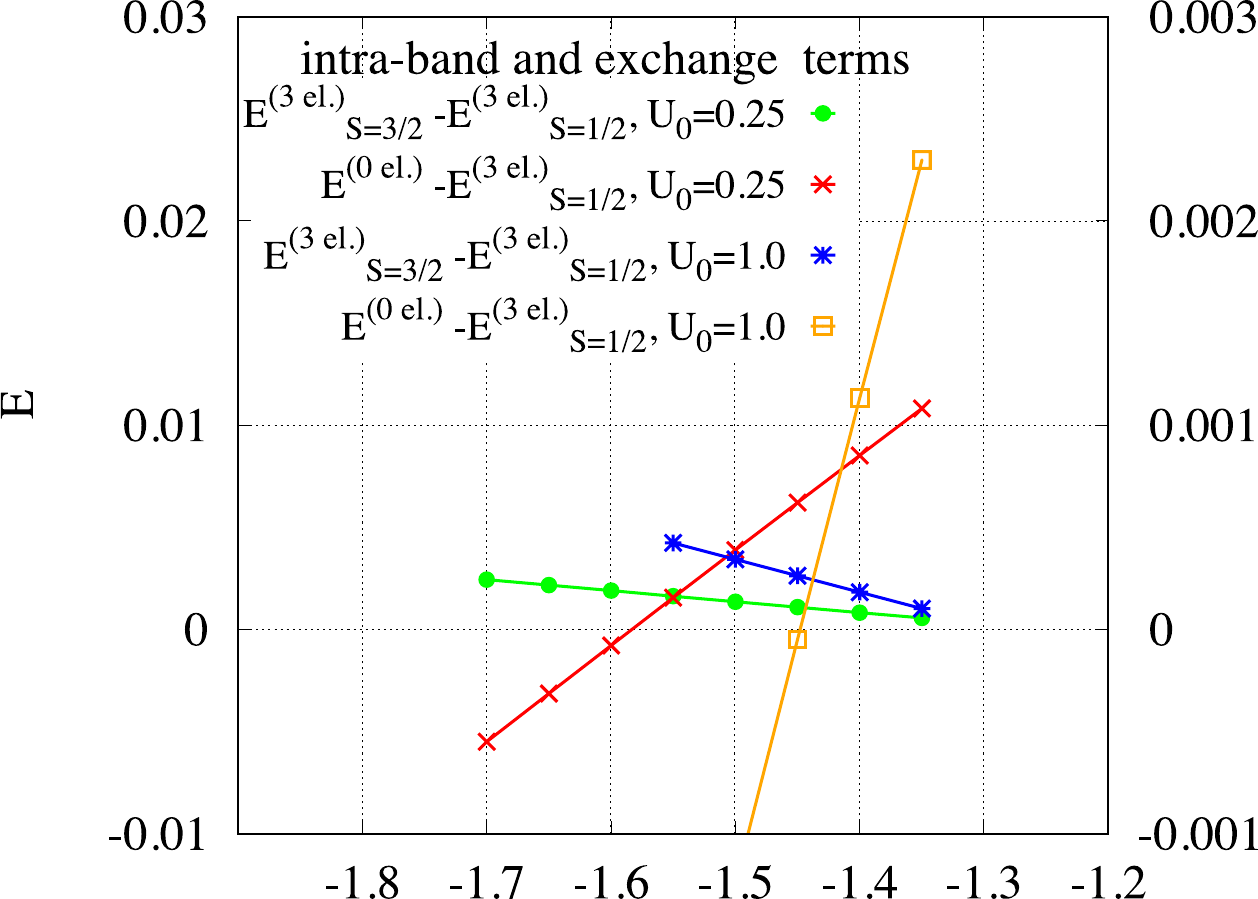}}
        \caption{(a) Effective spin-spin interaction and $U_{eff.} + 2 V_{eff.}$ interaction from intra-band terms only for the system depicted in the figure \ref{fig:system2_scheme} with the modified potentials and enlarged Kekule hoppings (see eq. \ref{eq:u_m_system3}) beyond the critical $\alpha=-1$ value. Left scale corresponds to charge-charge interaction, right scale corresponds to the spin-spin interaction. (b) Energy difference between $S=3/2$ and $S=1/2$ states with three electrons in the flat band; and energy difference between the state with empty flat band and the aforementioned $S=1/2$ state. Results obtained after exact diagonalization of the effective Hamiltonian \ref{eq:eff_ham_fin} taking into account exchange terms. The system is the same as in the figure (a). }
   \label{fig:system3_beyond_limit}
\end{figure}

\subsection{\label{sec:mapping_frustr}Frustrated spin-spin interaction beyond the edge of the parameters' region accessible for QMC}

Now we switch to $\alpha$ values outside of the limit accessible to QMC simulations.
Results for the effective spin-spin interaction $J^{(\Sigma)}_{eff.}$ obtained from intra-band terms at $\alpha<-\sqrt{1/10}$ are shown in 
Fig.~\ref{fig:system2_beyond_limit}.  As one can see, the spin-spin interaction becomes positive, thus the whole spin system in the flat band becomes frustrated. There is one peculiarity though. 
Large negative values of  $\alpha$ lead to large negative effective charge-charge interaction, which is evident from Fig.~\ref{fig:system2_analytical}. Eventually, it leads to the change of the ground state. If $\alpha$ is far enough from $-\sqrt{1/10}$, the ground state does not belong any more to the subspace with three electrons in the flat band. Instead, it belongs to the subspaces, where the number of electrons is either zero or six (empty or completely filled flat band).  We already saw this effect in the fitting procedure for the QMC data \ref{eq:fitting_curve2}, where we needed to take into account this state as a first excitation beyond the localized $S=3/2$ and $S=1/2$ states. 
The corresponding effective interaction $U_{eff.} + 2 V_{eff.}$ is shown in   Fig.~\ref{fig:system2_beyond_limit}.  If this interaction is  negative, the ground state becomes charged and we loose localization.  
However, it happens after $J^{(\Sigma)}_{eff.}$  becomes  positive, thus there is a  narrow region of $\alpha$
 values, where the electrons are still localized and simultaneously frustrated. In this  case,  the ground state has $S=1/2$ and is four-fold degenerate (two degenerate $S=1/2$ states with additional $M_z=\pm 1/2$ degeneracy). 

The energy scale in   Fig.~\ref{fig:system2_beyond_limit} is quite small. We can change this scale by changing the interaction matrix $U$. Example calculations are shown in Fig.~\ref{fig:system3_intra-band}.  The geometry of this system is the same as the one shown in Fig.~\ref{fig:system2_scheme}, but the larger Kekule hoppings are now equal to 1.5. $U$ matrix for the interacting dimers also differs. It is now defined as
\begin{eqnarray}
    U= U_0 \begin{pmatrix}
1 & \alpha \\
\alpha & 1  
\end{pmatrix}
  \label{eq:u_m_system3}
\end{eqnarray}
As one can see, the scale of the effective spin-spin interaction is now ten times larger and the frustrated ground state of localized spins exists in a much broader interval of $\alpha$ values. 

Finally, we check that the effect survives the introduction of exchange terms in the effective Hamiltonian.  The corresponding calculation is shown in 
Fig.\ref{fig:system3_inter-band}. Results are shown for $U_0=0.25$ and $U=1.0$, the values, where the results of the exact diagonalization of the approximate effective Hamiltonian were essentially indistinguishable from the exact  ones obtained using QMC. We compute the energies of $S=1/2$ and $S=3/2$ states in the subspace with three electrons in the flat band (denoted as $E^{(3 el.)}_{S=1/2}$ and  $E^{(3 el.)}_{S=2/2}$ in Fig.~\ref{fig:system3_inter-band}) and the energy of the empty flat band (denoted as $E^{(0 el.)}$). $E^{(3 el.)}_{S=3/2}>E^{(3 el.)}_{S=1/2}$ for both $U_0=0.25$ and $U=1.0$ for all $\alpha$ values used in the computation. This means that effective spin-spin interaction within the flat band is still frustrating even after the inclusion of the exchange terms for $\alpha<\alpha_1$, where $\alpha_1$ is some critical value. In addition to that $E^{(0 el.)}>E^{(3 el.)}_{S=1/2}$ for $\alpha>\alpha_0$ with $\alpha_0$ dependent on $U_0$. Combining these two inequalities, we conclude that there is an interval $\alpha_0<\alpha<\alpha_1$, where the electrons in the flat band are localized and frustrated. 

Even if there are some doubts about the approximations made during the derivation of the effective Hamiltonian \ref{eq:eff_ham_fin}, the limit $U_0/m \rightarrow 0$ stabilizes the frustrated state in any case. The reason is the diminishing role of the exchange terms in \ref{eq:eff_ham_fin} in the limit of a large gap. It means  that the intra-band results shown in the figure \ref{fig:system3_intra-band} eventually become exact, after appropriate rescaling due to decreasing $U_0$ and we still obtain frustrated localized spins within the flat band in a certain interval of $\alpha$ values.

\section*{\label{sec:Conclusion}Conclusion}

We constructed a two-dimensional lattice model with a flat band generated by adatoms placed on top of  a hexagonal  lattice.   An approximate effective Hamiltonian is constructed for the flat band assuming a large gap between conduction and valence bands. Our  projection  algorithm is verified using exact QMC data for the sign problem free system.  Subsequently, this verified projection algorithm is used to construct a lattice model, featuring a trimer of localized electrons in a flat band, interacting via frustrating antiferromagnetic interaction.  This lattice model is formulated on bipartite lattice and preserves particle-hole and SU(2) spin symmetry. However, in order to generate frustration in the effective Hamiltonian in the flat band, we need the combination of repulsive Hubbard interaction and large attractive long-range charge-charge interactions. This combination leads to the presence of both positive and negative eigenvalues in the charge-charge interaction matrix, thus making this model inaccessible   to  usual QMC simulations. 

We  hence  provide an explicit mapping between two classes of quantum systems, both of them  being  problematic for QMC calculations in the  sense  that  
they  generically  lead  to  a  negative  sign problem.  However, if  algorithmic  progress   is  achieved  for   systems with dominant long-range charge-charge interaction, 
our  mapping will immediately provide a solution for the sign problem for frustrated  spin  systems. 

As a final note, we would like to comment on the possibility to simulate systems where long-range interaction dominates over Hubbard interaction using the same trick with flat band projection. The question is if we can generate a sign problem free lattice model accessible to determinantal QMC, and featuring a flat band, where e.g. nearest-neighbour charge-charge interaction dominates over Hubbard repulsion.  As shown in the Appendix \ref{AppendixA}, the answer is negative. In order to make such simulations feasible, we again need to cross the point in the parameter space, where at least one eigenvalue of the interaction matrix  \ref{eq:int_hamiltonian} changes sign.  Thus such simulations are again unfeasible with present-day QMC algorithms.

More  generally,  flat  bands   provide  an  extremely  rich  playground  to investigate  correlation induced  phenomena,  a modern  example  being 
twisted  bilayer  graphene \cite{Cao18}.  While  single  electron hoping   is  suppressed  interactions in the particle-hole  channel,  as   described  here, 
as  well as  in the particle-particle channel  can  be  manipulated  by  tailoring  the    form of  the Wannier  functions  of the  flat  band  \cite{Peri20,Hofmann23}. 
Frustrated spin  systems, considered in the paper are not realistic due to the peculiar setup we used for the electron-electron interaction. However, the projection algorithm described in the paper can be used to model  experimentally relevant systems, where  frustration in the flat band,  formed by adatoms can be achieved through different mechanisms.  In   particular   if  we  relax  the  particle-hole symmetry condition  in the free Hamiltonian, $\hat H_K$,  then  the  Coulomb  interaction    
can  lead  to  frustrated effective spin-spin interaction in the flat band.   This  route  to design  frustrated  spin systems  is  left  for  future  research.

\begin{acknowledgments}

We thank E. Huffman for fruitful discussions. 
MU  thanks  the  DFG   for financial support  under the projects UL444/2-1. 
FFA   acknowledges financial support from the DFG through the W\"urzburg-Dresden Cluster of Excellence on Complexity and Topology in Quantum 
Matter - \textit{ct.qmat} (EXC 2147, Project No.\ 390858490)   as  well as  the SFB 1170 on Topological and Correlated Electronics at Surfaces and Interfaces (Project No.\  258499086).
The authors gratefully acknowledge the scientific support and HPC resources provided by the Erlangen National High Performance Computing Center (NHR@FAU) of the Friedrich-Alexander-Universität Erlangen-Nürnberg (FAU) under NHR project 80069. NHR funding is provided by federal and Bavarian state authorities. NHR@FAU hardware is partially funded by the German Research Foundation (DFG) through grant 440719683.
\end{acknowledgments}

\bibliography{general_bib, new_bib,fassaad.bib}

\pagebreak
\clearpage
\newpage

\appendix
\counterwithin{figure}{section}

\section{\label{sec:AppendixA} Stability of zero eigenvalue of interaction matrix}
\label{AppendixA}

\begin{figure}[]
   \centering
   \subfigure[]
   {\label{fig:2adatoms_1_scheme}\includegraphics[width=0.3\textwidth , angle=0]{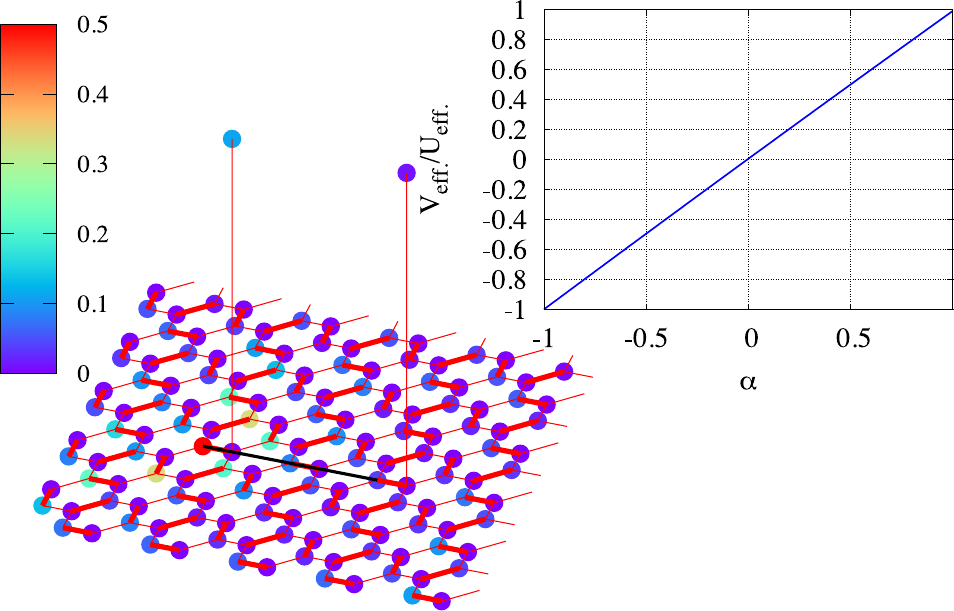}}
    \subfigure[]
   {\label{fig:2adatoms_2_scheme}\includegraphics[width=0.3\textwidth , angle=0]{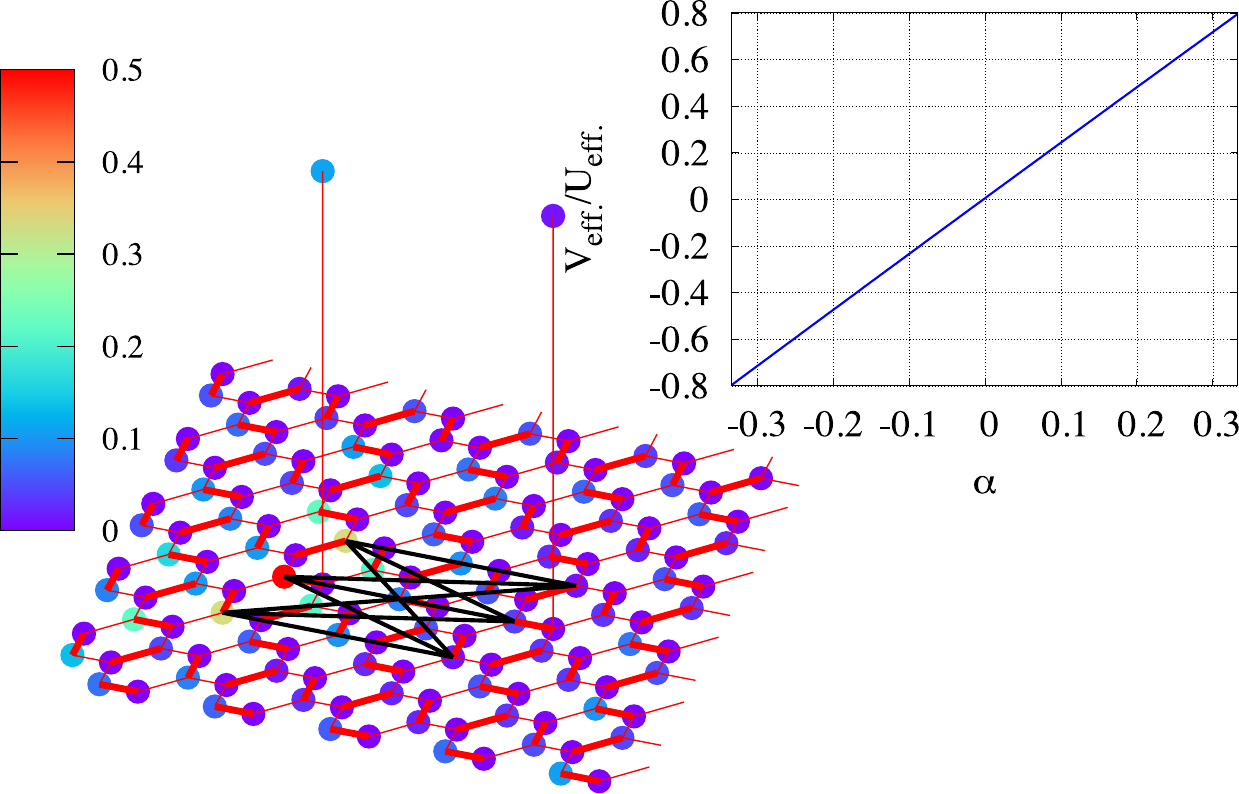}}
        \caption{Scheme of hoppings and interactions for system with two adatoms. Hoppings inside hexagonal lattice form Kekule pattern, bold lines correspond to $t=1.3$, thin lines correspond  to $t=1.0$. Hopping to adatom is equal to $t'=10.0$. Color scale corresponds to the modulus of the Wannier wavefunction, concentrated around one of the additional sites.  Figure (a) displays the setup when charge-charge interaction consists of one dimer \ref{eq:u_m_system_app1}. Figure (b) shows the situation when matrix of potential includes more connections between sites, where Wannier wavefunctions for flat band states have the largest absolute values \ref{eq:u_m_system_app2}. Insets: $\alpha$-dependence of the ratio of effective nearest-neighbour interaction  \ref{eq:V_eff_def} to effective Hubbard interaction  \ref{eq:U_eff_def} if only intra-band terms are taken into account. }
   \label{fig:2adatoms_schemes}
\end{figure}

 In this appendix we study the possibility to create a flat band system with nearest-neighbour interaction larger than the Hubbard interaction. Minimal system, where this effect can be studied consists of two adatoms, creating two zero modes in non-interaction Hamiltonian, as depicted in the schemes \textcolor{red}{A.}\ref{fig:2adatoms_1_scheme} and \textcolor{red}{A.}\ref{fig:2adatoms_2_scheme}. In the first scheme, matrix of interactions $U$ consists of only one dimer and is defined as 
 \begin{eqnarray}
    U= U_0 \begin{pmatrix}
1 & \alpha \\
\alpha & 1  
\end{pmatrix}.
  \label{eq:u_m_system_app1}
\end{eqnarray}
This dimer connects two equivalent sites near the adatoms. Interaction matrix has zero mode if $\alpha=\pm 1$, and this zero mode is preserved in the effective interactions $V_{eff.}$ and $U_{eff.}$ computed from intra-band terms \ref{eq:U_eff_def}, \ref{eq:V_eff_def}. The latter is evident from the inset on the figure \textcolor{red}{A.}\ref{fig:2adatoms_1_scheme}, where $V_{eff.}=-U_{eff.}$ at $\alpha=-1.0$.

 \begin{figure}[]
   \centering
     \subfigure[]{\includegraphics[width=0.3\textwidth , angle=270]{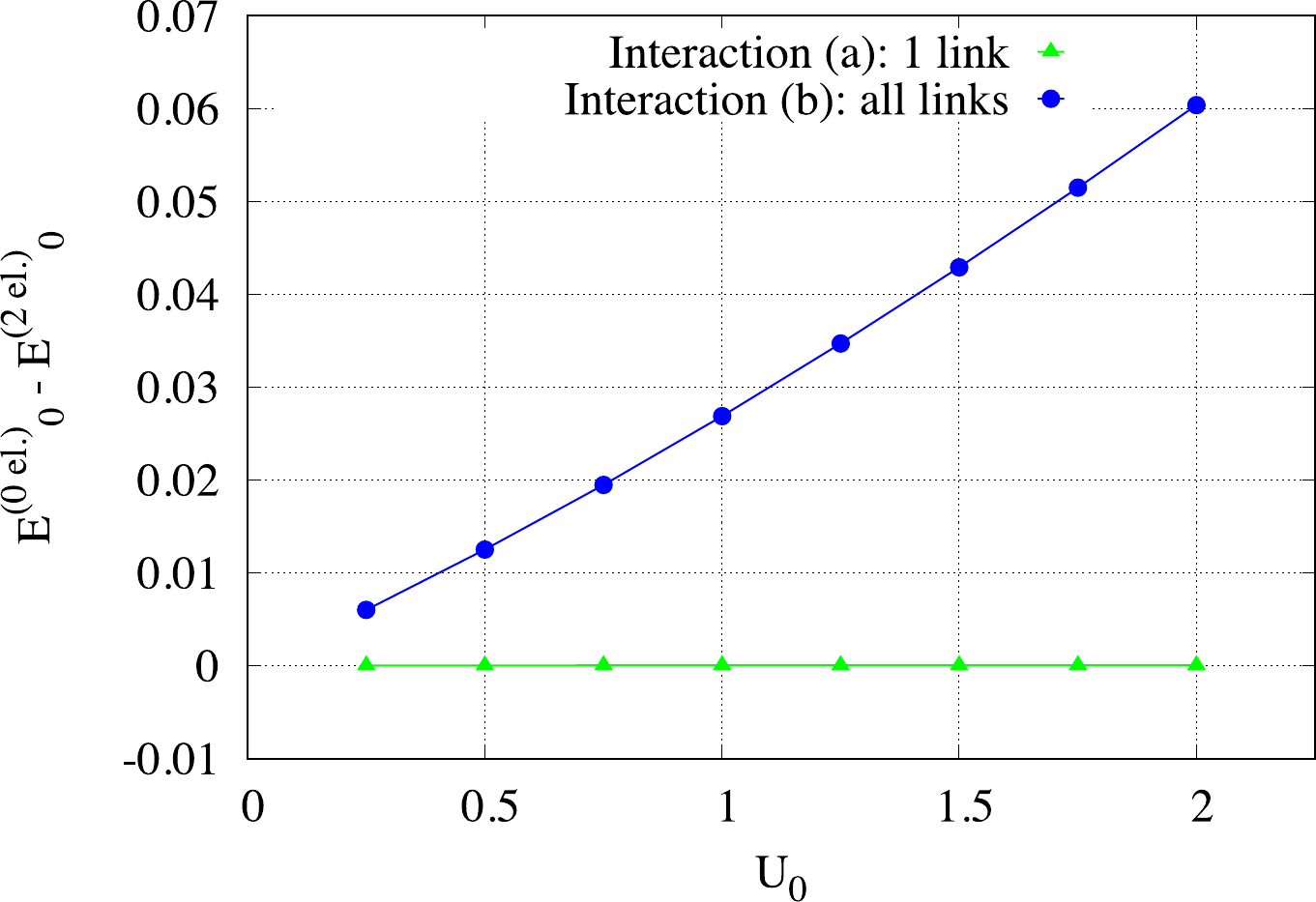}}
        \caption{Energy difference between the level, where the flat band is empty and the lowest level in  the subspace with two electrons in the flat band. The latter level corresponds in both cases to the degenerate state of localized spins with $S=1$ and $S=0$. Calculations were done for negative limiting values of $\alpha$: $\alpha=-1.0$ in (a) case and $\alpha=-1/3$ in (b) case. Corresponding systems are shown in the figures \textcolor{red}{A.}\ref{fig:2adatoms_1_scheme} and \textcolor{red}{A.}\ref{fig:2adatoms_2_scheme}.}
   \label{fig:2adatoms_energies}
\end{figure}

In the second scheme, the interaction Hamiltonian $H_U$ includes three sites with the largest electronic density of the corresponding Wannier function near each adatom. If the first three charges correspond to the sites near one adatom, and the last three charges correspond to the sites near another adatom, the matrix of interactions looks like: 
 \begin{eqnarray}
    U= U_0 \begin{pmatrix}
1 & 0      & 0 & \alpha & \alpha & \alpha  \\
0 & 1      & 0 & \alpha & \alpha & \alpha  \\
0 & 0      & 1 & \alpha & \alpha & \alpha  \\
\alpha & \alpha & \alpha & 1      & 0 & 0  \\
\alpha & \alpha & \alpha & 0      & 1 & 0 \\
\alpha & \alpha & \alpha & 0      & 0 & 1  
\end{pmatrix}.
  \label{eq:u_m_system_app2}
\end{eqnarray}
As displayed in the figure  \textcolor{red}{A.}\ref{fig:2adatoms_2_scheme}, the matrix of interactions includes all possible interactions between sites locates near different adatoms, but does not include  non-local interactions between sites located near the same adatom. $U$ matrix \ref{eq:u_m_system_app2} has zero eigenvalue if $\alpha=\pm 1/3$. However, the zero mode is not transferred to the effective interactions in this case:  it is evident from the inset in the figure \textcolor{red}{A}\ref{fig:2adatoms_2_scheme}: $V_{eff.} \neq -U_{eff.}$ at $\alpha=-1/3$, though absolute value of $V_{eff.}$ grows linearly with increasing $|\alpha|$.

The following broad statement can be formulated: we can write down a general interaction matrix and try to optimize it within the constraint that it has no negative eigenvalues. Maximum what we can get from intra-band terms only is the situation when $|V_{eff.}| = U_{eff.}$, and it appears only when there is a zero mode in the interaction matrix $U$. The first system (fig.  \textcolor{red}{A.}\ref{fig:2adatoms_1_scheme}) gives an example of such situation. In all other cases, the constraint leads to $|V_{eff.}| < U_{eff.}$ as it is for the second system (fig.  \textcolor{red}{A.}\ref{fig:2adatoms_2_scheme}). It also shows that the presence of the zero mode in $U$ matrix does not always lead to $|V_{eff.}| = U_{eff.}$.

The only remaining possibility is that the exchange terms can drive the effective potentials through this border ($|V_{eff.}| = U_{eff.}$).  In order to explore it, we diagonalize the full effective Hamiltonian \ref{eq:eff_ham_fin} and compare the energy of the state with empty flat band and the energy of the lowest state with two electrons in the flat band. Results for both systems are shown in the figure \ref{fig:2adatoms_energies}. We always choose the smallest negative $\alpha$ possible within the constraint of non-negative eigenvalues of the interaction matrix $U$.

For the first system, the two states in question remain degenerate independently from the interaction strength $U_0$. It means that $V_{eff.} = -U_{eff.}$ even after the inclusion of the exchange terms. Thus, the initial zero mode is stable not only with respect to the intra-band terms, but also with respect to the exchange terms.

For the second system, the energy difference is dependent on the interaction strength $U_0$, but it unfortunately evolves in the opposite direction to the desired one. Increasing $|V_{eff.}|$ with $V_{eff.}<0$ should lead to decreasing difference $E^{0 el.}_0 - E^{2 el.}_0$ so that the state with empty flat band becomes the ground state. The plot shows exactly opposite evolution.  

Thus we can conclude that there is no possibility to have  $|V_{eff.}|> U_{eff.}$ within the constraint of non-negative eigenvalues of the interaction matrix $U$. There are only two scenarios: 1)  $|V_{eff.}| = U_{eff.}$ in intra-band Hamiltonian (can only happen when $U$ has zero mode), but this situation remains unchanged after the inclusion of the exchange terms; 2) $|V_{eff.}| < U_{eff.}$ in intra-band Hamiltonian, but the exchange terms only reduce the $|V_{eff.}| / U_{eff.}$ ratio.

\end{document}